 \documentclass{emulateapj}

\usepackage{color}

\shorttitle{Low--redshift analogs of sub--mm galaxies}
\shortauthors{Oteo et al.}

\usepackage{natbib}

\begin{document}


\title{Low-redshift analogs of submm galaxies: a diverse population}


\author{I.~Oteo\altaffilmark{1,2},
I.~Smail\altaffilmark{3,4},
T.~M.~Hughes\altaffilmark{5,6}, 
L.~Dunne\altaffilmark{7,1}, 
R.~J.~Ivison\altaffilmark{2,1},
Z-Y.~Zhang\altaffilmark{1,2}, 
D.~Riechers\altaffilmark{8},
A.~Cooray\altaffilmark{9},
N.~Bourne\altaffilmark{1},
P.~van der Werf \altaffilmark{10},
D.~L. Clements\altaffilmark{11},
M.~J.~Micha{\l}owski\altaffilmark{12},
H. Dannerbauer\altaffilmark{13,14},
L. Wang\altaffilmark{15,16}
}

\affil{$^1$Institute for Astronomy, University of Edinburgh, Royal Observatory, Blackford Hill, Edinburgh EH9 3HJ UK}
\affil{$^2$European Southern Observatory, Karl-Schwarzschild-Str. 2, 85748 Garching, Germany}
\affil{$^3$Centre for Extragalactic Astronomy, Department of Physics, Durham University, South Road, Durham DH1 3LE UK}
\affil{$^4$Institute for Computational Cosmology, Department of Physics, Durham University, South Road, Durham DH1 3LE, UK}
\affil{$^5$Instituto de F\'isica y Astronom\'ia, Universidad de Valpara\'iso, Avda. Gran Breta\~na 1111, Valpara\'iso, Chile}
\affil{$^6$CAS Key Laboratory for Researches in Galaxies and Cosmology, Center for Astrophysics, Department of Astronomy, University of Science and Technology of China, Chinese Academy of Sciences, Hefei, Anhui 230026, China}
\affil{$^7$School of Physics and Astronomy, Cardiff University, The Parade, Cardiff CF24 3AA, UK}
\affil{$^8$Cornell University, Space Sciences Building, Ithaca, NY 14853, USA}
\affil{$^9$Department of Physics and Astronomy, University of California, Irvine, CA 92697, USA}
\affil{$^{10}$Leiden Observatory, Leiden University, P.O. Box 9513, NL-2300 RA Leiden, The Netherlands}
\affil{$^{11}$Physics Department, Blackett Lab, Imperial College, Prince Consort Road, London SW7 2AZ, UK}
\affil{$^{12}$Astronomical Observatory Institute, Faculty of Physics, Adam Mickiewicz University, ul.~S{\l}oneczna 36, 60-286 Pozna{\'n}, Poland}
\affil{$^{13}$Instituto de Astrof\'isica de Canarias (IAC), E-38205 La Laguna, Tenerife, Spain}
\affil{$^{14}$Universidad de La Laguna, Dpto. Astrof\'isica, E-38206 La Laguna, Tenerife, Spain}
\affil{$^{15}$SRON Netherlands Institute for Space Research, Landleven 12, 9747 AD, Groningen, The Netherlands}
\affil{$^{16}$Kapteyn Astronomical Institute, University of Groningen, Postbus 800, 9700 AV, Groningen, The Netherlands}
\email{ivanoteogomez@gmail.com}


  
\begin{abstract}

We have combined the wide--area {\it Herschel}--ATLAS far--IR survey with spectroscopic redshifts from GAMA and SDSS to define a sample of 21 low--redshift ($z_{\rm spec} < 0.5$) analogs of submm galaxies (SMGs). These have been selected because their dust temperatures and total IR luminosities are similar to those for the classical high--redshift SMG population.  As well as presenting the sample, in this paper we report $^{12}$CO(2--1) and $^{12}$CO(1--0) observations of 16 low--redshift analogs of SMGs taken with the IRAM--30m telescope.  We have obtained that low--redshift analogs of SMGs represent a very diverse population, similar to what has been found for high--redshift SMGs.  A large variety in the molecular gas excitation or $^{12}$CO(2--1)/$^{12}$CO(1--0) line ratio is seen, meaning that extrapolations from $J \geq 2$ CO lines can result in very uncertain molecular gas mass determinations.  Our sources with $^{12}$CO(1--0) detections follow the dust--gas correlation found in previous work at different redshifts and luminosities.  The molecular gas mass of low--redshift SMGs has an average value of $M_{\rm H_2} \sim 1.6 \times 10^{10}\,M_\odot$ and will be consumed in $\sim 100 \, {\rm Myr}$ .  We also find a wide range of molecular gas fractions, with the highest values being compatible with those found in high--redshift SMGs with $^{12}$CO(1--0) detections, which are only the most luminous.  Low--redshift SMGs offer a unique opportunity to study the properties of extreme star formation in a detail not possible at higher redshifts.

\end{abstract}


%

\section{Introduction}\label{intro}

Twenty years ago, the first deep, single--dish observations of the sky at submm wavelengths revealed a population of submm galaxies \citep[SMGs,][]{Smail1997ApJ...490L...5S,Barger1998Natur.394..248B,Hughes1998Natur.394..241H}, which, since then, have revolutionized our understanding of the formation and evolution of galaxies.  Soon after their discovery, it was reported that SMGs had a median redshift of $z \sim 2.5$ \citep{Chapman2003Natur.422..695C,Chapman2005ApJ...622..772C,Simpson2014ApJ...788..125S,Danielson2017ApJ...840...78D}.  The importance of SMGs for galaxy evolution was later highlighted by {\it Spitzer} and {\it Herschel} studies, which agreed that SMGs host up to 30\% the star formation occurring at $z \sim 2-3$ \citep{Chapman2005ApJ...622..772C,Magnelli2011A&A...528A..35M}. SMGs have been proposed to be scaled--up analogs of local ULIRGs which are dominated by merger--induced starbursts \citep{Tacconi2008ApJ...680..246T}.  However, SMGs have lower dust temperature at a fixed IR luminosity \citep{Hwang2010MNRAS.409...75H} and seem to have higher molecular gas fractions and more intense star formation \citep{Carilli2013ARA&A..51..105C}.  

Despite representing a key population for galaxy evolution, detail studies of the classical SMG population have been limited by a combination of their high redshift (meaning that they have small sizes which require very high--spatial resolution -- \citealt{Bussmann2015ApJ...812...43B,Simpson2015ApJ...807..128S,Hodge2016ApJ...833..103H}) and dusty nature (meaning that they become extremely faint at rest--frame UV and optical wavelengths and hindering the study of their stellar emission).

\begin{figure*}
\centering
\includegraphics[width=0.9\textwidth]{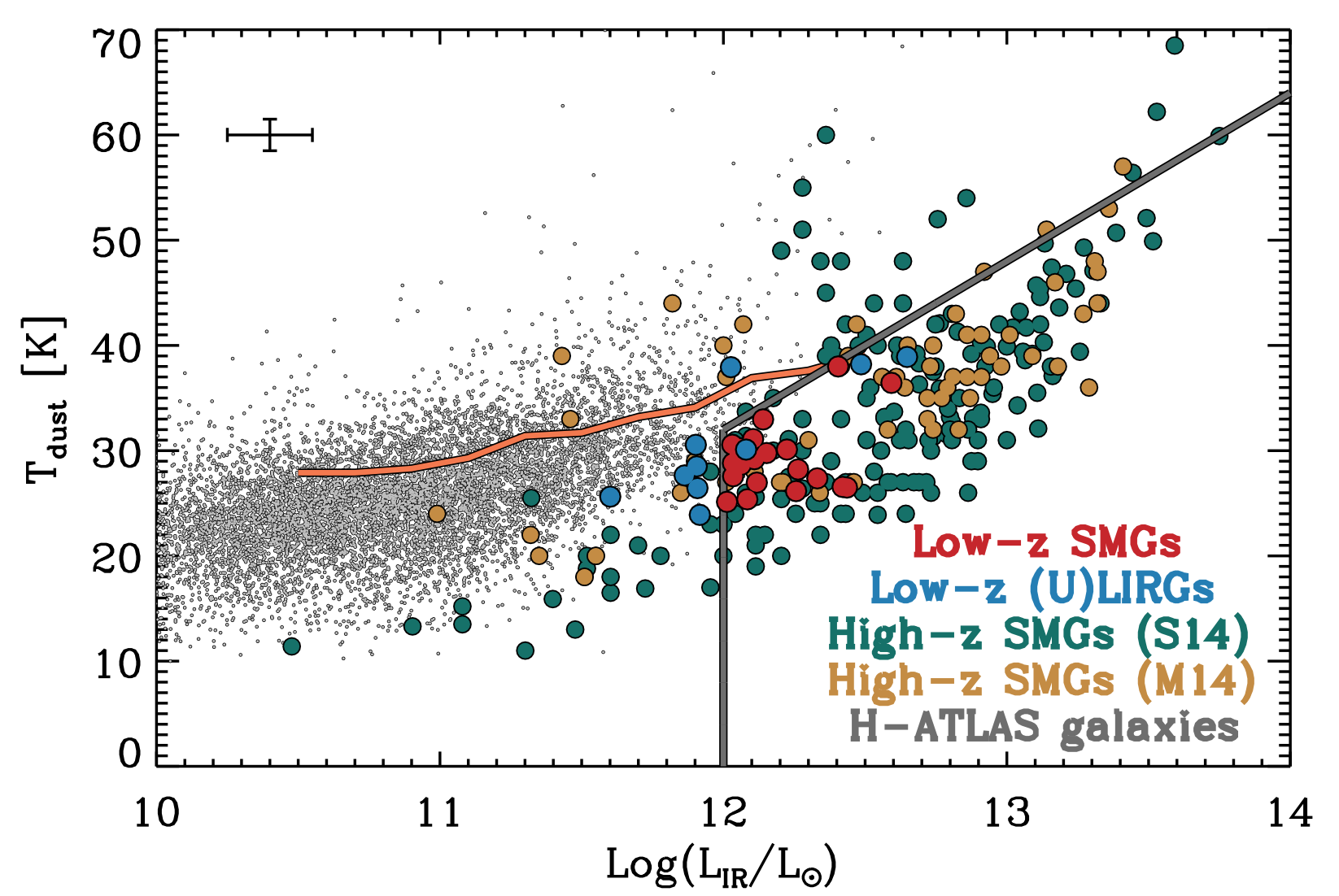} 
\caption{Dust temperature ($T_{\rm dust}$) as a function of the total IR luminosity ($L_{\rm IR}$) for all galaxies in {\it H}--ATLAS with secure spectroscopic redshifts and reliable FIR--to--optical association. The dust temperatures have been obtained by fitting optically thin dust emission models and assuming a fixed dust emissivity of $\beta = 1.5$. The typical error bar is shown in the upper--left corner. Our low--redshift SMGs are selected because their dust temperatures and total IR luminosities are similar to those found for the classical SMG population at high redshift (selection window shown with the black solid lines). The dust temperature and total IR luminosity for the high--redshift SMGs have been taken from \cite{Magnelli2012A&A...539A.155M} and \cite{Swinbank2014MNRAS.438.1267S}.  For comparison, we show the location of the low--redshift (U)LIRGs studied in \cite{Magdis2013A&A...558A.136M}, three of which would have been selected as low--redshift SMGs.  Furthermore, we also include the $T_{\rm dust} - L_{\rm IR}$ relation derived from IR--bright galaxies selected in deep and small--area {\it Herschel} surveys \citep{Symeonidis2013MNRAS.431.2317S}.
              }
\label{selection_low_redshift_SMGs_TD_LIR}
\end{figure*}

The main advantage of using low--redshift analogs is that they allow studies in a level of detail (with much better spatial resolution and sensitivity) not possible at high redshift even with the most powerful instrumentation currently available.  \cite{Heckman2005ApJ...619L..35H} used GALEX in combination with SDSS to create a sample of nearby ($z_{\rm spec} < 0.3$) galaxies (excluding AGNs) with far--UV luminosities and far--UV surface brightness chosen to overlap with the luminosity range of high--redshift Lyman--break galaxies \citep[see also][]{Hoopes2007ApJS..173..441H,Greis2016MNRAS.459.2591G,Contursi2017arXiv170604107C}.  The aim was to use low--redshift Lyman-break analogs to provide an opportunity to explore the physical processes occurring in typical star--forming galaxies at much higher redshifts \citep{Overzier2010ApJ...710..979O}.  A similar approach has been used for another of the classical high--redshift populations: Lyman--$\alpha$ emitters \citep{Deharveng2008ApJ...680.1072D,Cowie2010ApJ...711..928C,Cowie2011ApJ...738..136C,Oteo2011ApJ...735L..15O,Oteo2012ApJ...751..139O,Hayes2013ApJ...765L..27H,Atek2014A&A...561A..89A}.  Therefore, low--redshift analogs represent an important population for understanding and interpreting the properties of galaxies at high redshift.  With this in mind, we identify and use in this work a sample of 21 low--redshift SMGs as a reference to understand the important population of high--redshift SMGs. It should be pointed out that our sources are not only the best analogues in the low--redshift Universe of classical high--redshift SMGs, but also a unique population to study the ISM of dusty galaxies at low--redshift, filling the gap between local ($z < 0.05$) and $z > 1$ studies, and being complementary in luminosity, redshift, or dust temperature than sources in other surveys \citep{Combes2011A&A...528A.124C,Combes2013A&A...550A..41C,Villanueva2017arXiv170509826V}.

The paper is structured as follows: In \S \ref{section_source_selection} we present the sample selection.  Then \S \ref{section_IRAM_observations} describes the IRAM--30m observations used in this paper.  The main results of the paper are shown in \S \ref{results_of_the_paper}, including the analysis of the CO line ratios, scaling relations, or the molecular gas fraction.  Finally, the main conclusions of the paper are shown in \S \ref{section_conclusions_paper}. The total IR luminosities ($L_{\rm IR}$) reported in this work refer to the integrated luminosities between rest--frame 8 and $1000\,{\rm \mu m}$. Throughout this paper, the reported SFRs are derived from $L_{\rm IR}$ by assuming a Salpeter IMF (the same is used for the stellar mass estimation) and the classical \cite{Kennicutt1998ARA&A..36..189K} calibration. We assume a flat Universe with $(\Omega_m, \Omega_\Lambda, h_0)=(0.3, 0.7, 0.7)$.

\begin{table*}
\caption{\label{table_full_sample}Properties of our sample of low-$z$ SMGs}
\centering
\begin{tabular}{cccccccccccccccccc}
\hline
IAU Name & Source & $z_{\rm spec}$ & $S_{\rm 100 \mu m}$ & $S_{\rm 160 \mu m}$ &	$S_{\rm 250 \mu m}$ & $S_{\rm 350 \mu m}$	&	$S_{\rm 500 \mu m}$	& $M_{\rm star}$\tablenotemark{(a)} & $M_{\rm dust}$\tablenotemark{(b)} \\
& & & [mJy] & [mJy] &	[mJy] & [mJy]	&	[mJy]	& [$\times 10^{11} \, M_\odot$] & [$\times 10^{8} \, M_\odot$]\\
\hline\hline
HATLAS\,J085623.6+000628	&	G09.DR1.191		& $ 0.375 $	& $ 223 \pm 41 $ 	& $ 203 \pm 44 $	& $ 178 \pm 7 $ 	& $ 101 \pm 8 $	& $ 27 \pm 9 $		& $\sim 1.4 $	&		$\sim 2.42$ \\
HATLAS\,J090208.2+000935 	&	G09.DR1.333		& $ 0.418 $ 	& $ 170 \pm 27 $ 	& $ 206 \pm 32 $	& $ 140 \pm 7 $ 	& $ 65 \pm 8 $		& $ 40 \pm 9 $ 		& $\sim 0.5 $	&		$\sim 1.45$ \\
HATLAS\,J091340.1+010056	&	G09.DR1.344		& $ 0.422 $	& $ 178 \pm 39 $ 	& $ 256 \pm 47 $	& $ 122 \pm 6 $ 	& $ 62 \pm 8 $ 		& $ 23 \pm 8 $ 		&  $\sim 2.3$	&		$\sim 1.02$ \\
HATLAS\,J085429.5$-$003828	&	G09.DR1.350		& $ 0.392 $ 	& $ 162 \pm 38 $	& $ 181 \pm 46 $	& $ 129 \pm 7 $	& $ 57 \pm 8 $		& $ 9	 \pm 9 $  		& $\sim 3.0 $	&		$\sim 1.56$ \\
HATLAS\,J083733.8+000055	&	G09.DR1.370		& $ 0.382 $	& $ 174 \pm 39 $      	& $ 211 \pm 45 $	& $ 120 \pm 7 $      	& $ 59 \pm 8 $		& $ 11 \pm 9 $  		& $\sim 3.9 $	&		$\sim 1.31$ \\
HATLAS\,J085307.2$-$004954	&	G09.DR1.444		& $ 0.451 $     	& $ 202 \pm 44 $	& $ 172 \pm 46 $      	& $ 114 \pm 7 $  	& $ 71 \pm 8 $  	& $ 23 \pm 9 $ 		& $\sim 0.4 $	&		$\sim 1.59$ \\
HATLAS\,J085157.4$-$001230	&	G09.DR1.575		& $ 0.461 $	& $ 188 \pm 39 $	& $ 254 \pm 48 $	& $ 117 \pm 7 $       	& $ 56 \pm 8 $		& $ 17 \pm 9 $ 		& $\sim 0.8 $	&		$\sim 1.36$ \\
HATLAS\,J084959.2$-$003400	&	G09.DR1.611		& $ 0.388 $	& $ 117 \pm 29	$	& $ 139 \pm 32 $	& $ 96 \pm 6 $		& $ 40 \pm 8 $		& $ 3 \pm 8 $		& $\sim 1.1 $	&		$\sim 1.14$ \\
HATLAS\,J091637.9+001155	&	G09.DR1.978		& $ 0.357 $ 	& $ 135 \pm 38 $	& $ 124 \pm 45 $	& $ 94 \pm 7 $		& $ 31 \pm 8 $ 		& $ 2 \pm 9 $		&  $\sim 2.2 $	&		$\sim 0.87$ \\
HATLAS\,J091409.0+005543	&	G09.DR1.1442		& $ 0.339 $	& $ 118 \pm 43 $	& $ 209 \pm 45 $	& $ 89 \pm 7 $		& $ 32 \pm 8 $ 		& $ 9 \pm 9 $ 		&  $\sim 1.3 $	&		$\sim 0.69$ \\
HATLAS\,J120626.6+000931	&	G12.DR1.242		& $ 0.413 $      	& $ 161 \pm 40 $	& $ 257 \pm 46 $	& $ 166 \pm 7 $       	& $ 83 \pm 8 $		& $ 24 \pm 9 $		&  $\sim 5.1 $	&		$\sim 2.45$ \\
HATLAS\,J120715.7$-$010017	&	G12.DR1.254		& $ 0.385 $       & $ 481 \pm 42 $	& $ 372 \pm 47 $	& $ 165 \pm 7 $        & $ 80 \pm 8 $        	& $ 28 \pm 9 $  	&  $\sim 0.2 $	&		$\sim 1.10$ \\
HATLAS\,J113653.9+000621	&	G12.DR1.287		& $ 0.407 $       & $ 142 \pm 39 $	& $ 206 \pm 48 $	& $ 150 \pm 7 $    	& $ 71 \pm 8 $ 		& $ 13 \pm 9 $		&  $\sim 1.4 $	&		$\sim 2.22$ \\
HATLAS\,J115414.6$-$004721	&	G12.DR1.291		& $ 0.450 $	& $ 198 \pm 41 $     	& $ 240 \pm 47 $	& $ 158 \pm 7 $	& $ 93 \pm 8 $        	& $ 44 \pm 9 $ 		&  $\sim 2.6 $	&		$\sim 2.79$ \\
HATLAS\,J115329.3$-$003453	&	G12.DR1.313		& $ 0.450 $	& $ 132 \pm 43 $      	& $194 \pm  47 $	& $ 149 \pm 7 $	& $ 89 \pm 8 $       	& $ 30 \pm 9 $		&  $\sim 3.6 $ 	&		$\sim 3.21$ \\ 
HATLAS\,J115414.3$-$015609	&	G12.DR1.425		& $ 0.410 $	& $ 160 \pm 41 $	& $ 252 \pm 46 $     	& $ 131 \pm 7 $	& $ 76 \pm 8 $         	& $ 21 \pm 9 $		&  $\sim 4.9 $	&		$\sim 1.84$ \\
HATLAS\,J114224.1$-$010659	&	G12.DR1.567		& $ 0.478 $	& $ 122 \pm 41 $     	& $ 188 \pm 47 $	& $ 120 \pm 7 $	& $ 60 \pm 8 $		&  $	8 \pm 9 $ 		&  $\sim 1.1 $ 	&		$\sim 1.97$ \\     
HATLAS\,J114018.1+004130	&	G12.DR1.762		& $ 0.450 $	& $ 139 \pm 40 $	& $ 165 \pm 52 $	& $ 119 \pm 7 $		& $ 62 \pm 8 $		& $ 21 \pm 9 $		&  $\sim 1.5 $	&		$\sim 1.86$ \\
HATLAS\,J144000.9+010740	&	G15.DR1.83		& $ 0.456 $	& $ 493 \pm 28 $      & $ 451 \pm 32 $      	& $ 246 \pm 6 $	& $ 117 \pm 7 $		& $ 36 \pm 8 $		&  $\sim 1.5 $	&		$\sim 2.55$ \\ 
HATLAS\,J144135.0+014559	&	G15.DR1.244		& $ 0.436 $	& $ 171 \pm 37 $	& $ 232 \pm 45 $ 	& $ 161 \pm 6 $	& $ 92 \pm 7 $       	& $ 31 \pm 8 $ 		&  $\sim 4.4 $	&		$\sim 2.78$ \\
HATLAS\,J143027.6$-$005614	&	G15.DR1.318 		& $ 0.318 $	& $ 194 \pm 39 $	& $ 216 \pm 44 $	& $ 138 \pm 6 $	& $ 58 \pm 7 $		& $ 15 \pm 8 $		&  $\sim 2.4 $	&		$\sim 1.17$ \\
\hline
\hline
\tablenotetext{1}{Estimated from the best--fitted MAGPHYS templates}
\tablenotetext{2}{Estimated from the same fits used to determine the dust temperature (see \S \ref{section_source_selection}) and using Equation 8 in \cite{Casey2012MNRAS.425.3094C}. The dust absorption factor is assumed to be $\kappa_{850} = 0.15 \, {\rm m^2 \, kg^{-1}}$.}
\end{tabular}
\end{table*}

\section{Source selection}\label{section_source_selection}
 
Our sample of low--redshift analogs of SMGs has been selected from the combination of {\it H}--ATLAS, the widest far--IR survey carried out with the {\it Herschel} Space Observatory \citep{Eales2010PASP..122..499E,Valiante2016MNRAS.462.3146V,Bourne2016MNRAS.462.1714B}, and photometry and spectroscopy from the GAMA optical survey \citep{Baldry2010MNRAS.404...86B,Driver2011MNRAS.413..971D,Liske2015MNRAS.452.2087L}.  As an previous step for the selection of low--redshift SMGs we have derived the total IR luminosity and dust temperature of all {\it H}-ATLAS galaxies with available spectroscopic redshift and robust association between the {\it Herschel} and the optical emission (matching probability higher than 80\%).  We point out that a higher matching probably has not been chosen to avoid biasing the sample against mergers (it is hard to achieve high matching probability if there are two galaxies very close to the {\it Herschel} position).  In order to determine the most reliable values of dust temperature, we have considered only those galaxies in {\it H}--ATLAS which are detected in all PACS and SPIRE bands. The dust temperature of each source has been derived by fitting optically thin dust emission models to its observed PACS and SPIRE flux densities assuming a fixed dust emissivity of $\beta = 1.5$, which is the average value found for high-redshift SMGs \citep[see for example][]{Magnelli2012A&A...539A.155M}. The total IR luminosity of each source has been derived by fitting its mid--IR (data taken from the WISE all--sky survey) to far--IR SED with a set of templates associated to different local and high--redshift starbursts, including M\,82, Mrk\,231, Arp\,220, the average template of high--redshift SMGs \citep{Swinbank2014MNRAS.438.1267S} or the Eyelash \citep{Swinbank2010Natur.464..733S}. Then the best--fit template for each source has been integrated between rest--frame 8 and $1000 \, {\rm \mu m}$.  The relation between dust temperature and total IR luminosity of all {\it H}--ATLAS sources is shown in Figure \ref{selection_low_redshift_SMGs_TD_LIR}.

Our sample of low--redshift analogues of SMGs is formed by all {\it H}--ATLAS sources in the GAMA fields (see below why we need to restrict to GAMA fields) with spectroscopic redshift $z_{\rm spec} < 0.5$ and whose dust temperature and total IR luminosities are similar to those for high--redshift SMGs.  We show the selection window for low--redshift SMGs in Figure \ref{selection_low_redshift_SMGs_TD_LIR}, which is defined as the region where more than 80\% of high--redshift SMGs are located.  In this way, our sample of low--redshift SMGs is selected to comprise the most luminous and coldest {\it Herschel} sources with $z_{\rm spec} < 0.5$. This is equivalent to imposing that the far--IR SEDs of high--redshift and low--redshift SMGs are similar.     

The spectroscopic redshift requirement means that we need to limit the low--redshift SMG selection to the {\it H}--ATLAS GAMA fields, since the spectroscopic information in the other two {\it H}--ATLAS fields, NGP and SGP, is very limited.  We note that the most significant source of incompleteness in our sample is the lack of spectroscopic redshifts for all {\it H}--ATLAS sources: there might be low--redshift SMGs in the {\it H}--ATLAS GAMA fields which are not included in our sample because there are not available spectroscopic redshifts for them (we recall that spectroscopic redshifts are needed for the CO observations, one of the aims of this project).  Therefore, we do not aim at building a complete sample of low--redshift SMGs, but instead a representative sample to study extreme star formation in a level of detail which is not possible in the high--redshift Universe.  

The current sample of low--redshift SMGs is formed by 21 sources, whose main properties are quoted in Table \ref{table_full_sample}.  The number of sources in our low--redshift SMG sample might increase in the future once more spectroscopic redshifts for {\it H}--ATLAS sources are available.  This will actually help to populate the region in the $T_{\rm dust} - L_{\rm IR}$ diagram associated to the lowest dust temperatures found for SMGs and also add sources in the most luminous end.

We note that our sample of low--redshift SMGs is dissimilar to most populations of IR--bright galaxies studied in the literature.  For example, our sources are more luminous and are at higher redshifts than sources in the VALES survey \citep{Villanueva2017arXiv170509826V}.  Despite their comparable IR luminosities, our low--redshift SMGs are colder than most ULIRGs studied so far at their same redshift \citep{Combes2011A&A...528A.124C}.  The most comparable sample of IR--bright galaxies to our low--redshift SMGs are the ULIRGs studied in \cite{Magdis2013A&A...558A.136M}, three of which would have been selected as low--redshift SMGs according to our criterion (see Figure \ref{selection_low_redshift_SMGs_TD_LIR}).  Therefore, in addition to study the properties of arguably the best low--redshift analogs of high--redshift SMGs, we are also explore a parameter space not studied in detail before with a relatively large sample of galaxies.

\begin{figure*}
\centering
\includegraphics[width=0.80\textwidth]{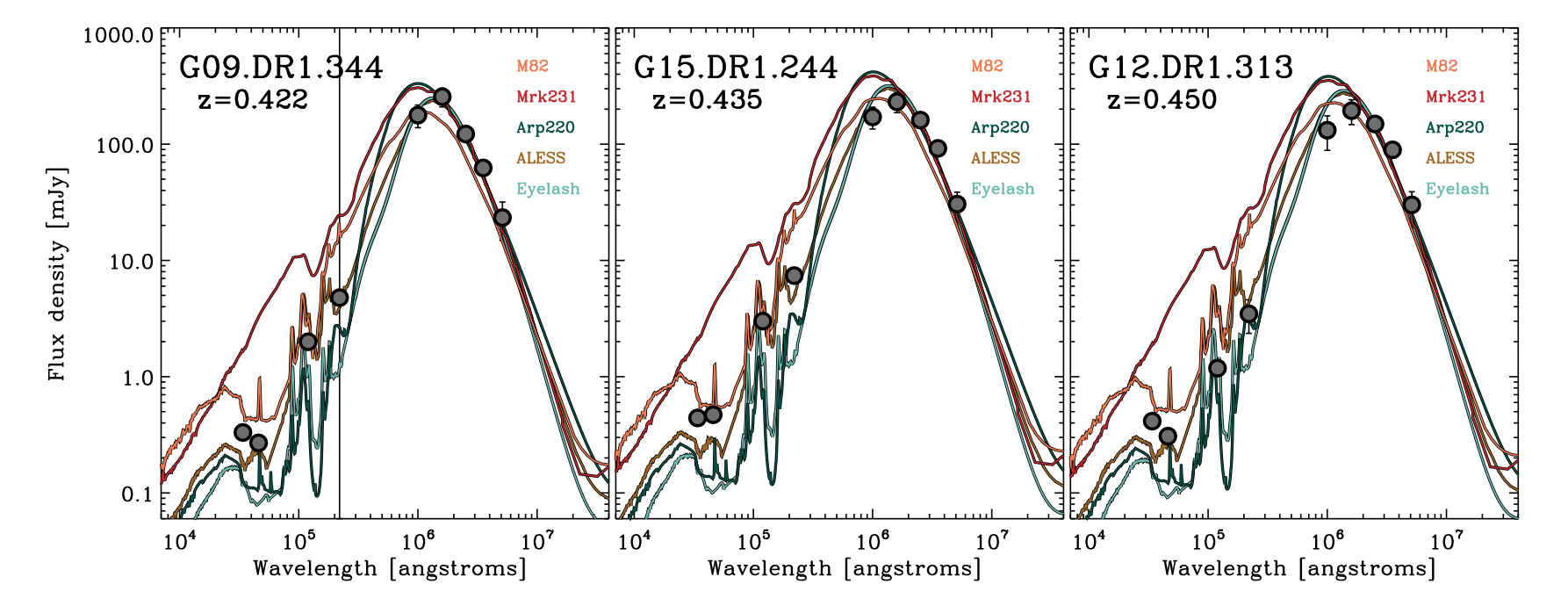}
\caption{Near--IR to far--IR SED of three of our low--redshift SMGs. We have included WISE and {\it Herschel} photometry along with a set of templates associated to known starbursts at low and high redshift, which have been fitted to the {\it Herschel} photometry only. The observed photometry of most of our low--redshift SMGs is well fitted by the ALESS template, which represents the average SED of the classical population of high--redshift SMGs \citep{Swinbank2014MNRAS.438.1267S}.  This is a natural result, as our low--redshift SMGs have been selected because they occupy the same region in the $T_{\rm dust} - L_{\rm IR}$ diagram as high--redshift SMGs and, consequently, are expected to have similar far--IR SEDs. Note that some low--redshift SMGs are even colder (the FIR SED peaks at longer wavelengths) than the average high--redshift SMG, see for example G12.DR10.313.  The mid--IR photometry of our sources is also compatible with that found for the average high--redshift SMG. The SEDs for all the galaxies in our sample of low--redshift SMGs can be found in the Appendix.
              }
\label{sed_low_redshift_SGMS_WISE}
\end{figure*}

Figure \ref{sed_low_redshift_SGMS_WISE} shows the SED of some of our low--redshift SMGs (the SEDs for the full sample are shown in the Appendix). The FIR SED of most low--redshift SMGs is very well represented by the ALESS template, which represents the average SED of the classical population of high--redshift SMGs \citep{Swinbank2014MNRAS.438.1267S}.  This is an expected result since this template represents the average SED of high--redshift SMGs and our galaxies are selected for having similar SEDs to high--redshift SMGs. We point out that some of our galaxies are even redder (colder) than predicted by the ALESS template, such as G12.DR1.242 or G12.DR1.287.  The mid--IR photometry of our sources is compatible with that seen in high--redshift SMGs, and the mid--IR colors do not suggest strong contribution of AGNs in our sample.

\begin{figure}
\centering
\includegraphics[width=0.45\textwidth]{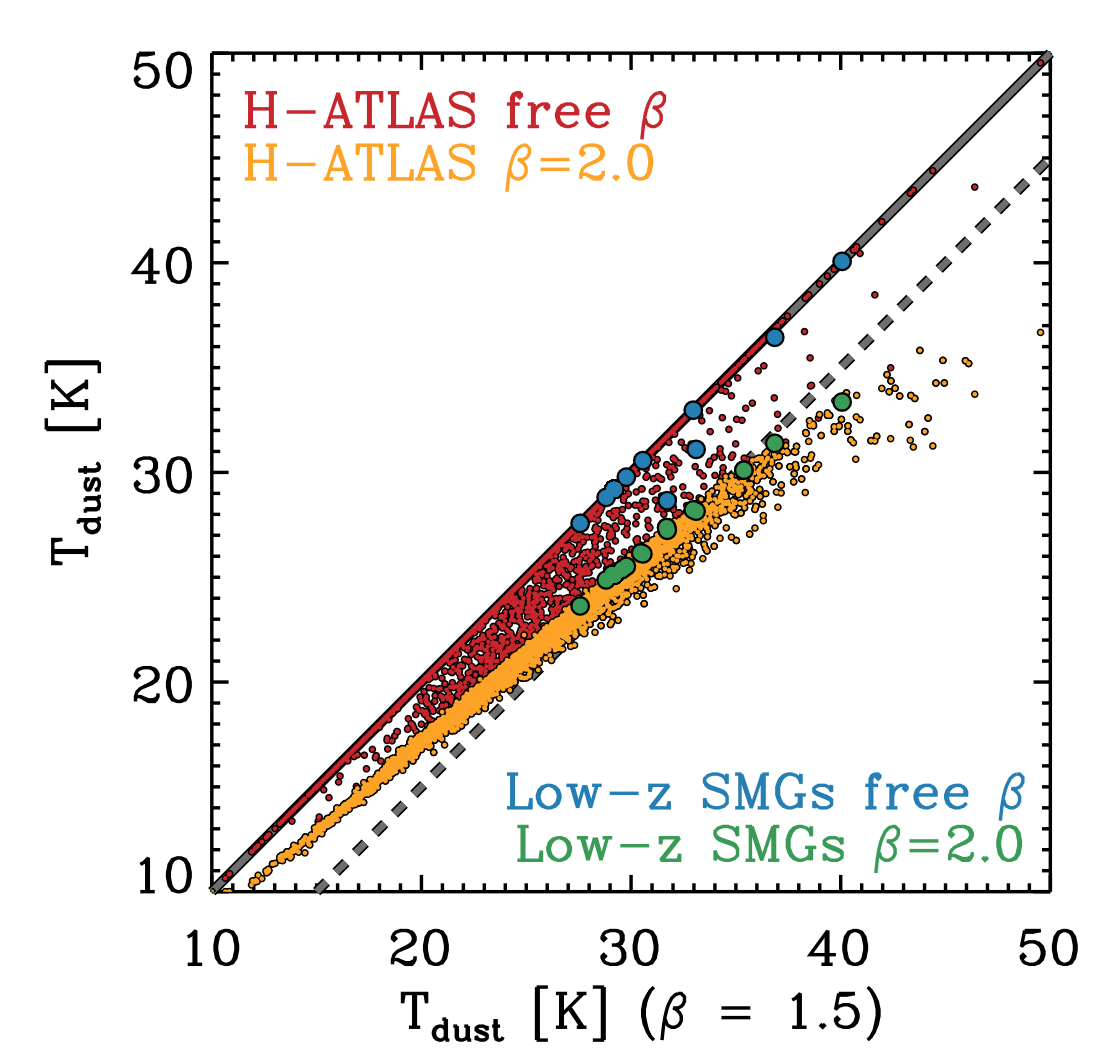} 
\caption{Dust temperature ($T_{\rm dust}$) for the full sample of {\it H}--ATLAS galaxies with good optical association and for our low--redshift SMGs.  The dust temperatures used in this work to select low--redshift SMGs have been obtained by assuming $\beta = 1.5$, and in this figure we compare them to the values found for $\beta = 2.0$ and when $\beta$ is left as a free parameter (with values allowed to range within $1.5 < \beta < 2.0$) in the FIR SED fitting with modified black--body functions.  We over--plot the one--to--one relation with a solid line and a deviation of $- 5\,{\rm K}$ with respect to the one--to--one relation.  We see that, for our low--redshift SMGs, the deviations in dust temperature when using different assumptions for $\beta$ can be up to $\sim 5 \, {\rm K}$.  That this uncertainty in not only present in this work, but also in most previous work studying dust temperature from {\it Herschel} data.  It is worth noting that when leaving $\beta$ as a free parameter, the preferred value of $\beta$ in our low--redshift SMGs is close to $\beta \sim 1.5$.
              }
\label{fig_beta_temperature_fig}
\end{figure}

Since the values of the dust temperature are involved in the selection of our low--redshift SMGs, we now discuss the uncertainties related to their determination.  In order to obtain the dust temperature of the full sample of {\it H}--ATLAS galaxies with spectroscopic redshifts (which our low--redshift SMGs are selected from), we have assumed a fixed dust emissivity, $\beta = 1.5$.  However there is a degeneracy between dust temperature and dust emissivity which introduces an uncertainty in their calculation.  Figure \ref{fig_beta_temperature_fig} shows the relation between the dust temperature of {\it H}--ATLAS galaxies and our low--redshift SMGs when assuming $\beta = 1.5$, $\beta = 2.0$, and when $\beta$ is left as a free parameter (with values allowed to vary within $1.5 < \beta < 2.0$).  We have considered this range of variability in $\beta$ because it is the same as the one used by \citep{Swinbank2014MNRAS.438.1267S} in their study of high--redshift SMGs.  Note that \cite{Magnelli2012A&A...539A.155M} used $\beta = 1.5$ in their study of high--redshift SMGs, in part because this choice is fully compatible with the $\beta$ values that they find when this parameter is left free in the FIR SED fits.  We see in Figure \ref{fig_beta_temperature_fig} that significant dust temperature variations can happen for the full population of {\it H}--ATLAS sources, and also for high--redshift SMGs.  In general, we see that higher $\beta$ means lower dust temperature, as expected.  The dust temperature of our low--redshift SMGs can vary up to $\sim 5 \, {\rm K}$ depending on the value of $\beta$.  This uncertainty in the dust temperature determination is present not only in our work, but in all previous work measuring dust temperatures from {\it Herschel} data.  

From the analysis above we conclude that the most significant uncertainty in the determination of dust temperature is the assumption of the dust emissivity $\beta$.  Therefore, it is key that we use the same $\beta$ assumptions that was used in the high--redshift SMG samples we compare to. In fact, we have used $\beta = 1.5$ as in \cite{Magnelli2012A&A...539A.155M}.  \cite{Swinbank2014MNRAS.438.1267S} used a free $\beta$ in the fits, with values ranging within $1.5 < \beta < 2.0$.  If we use this option, we obtain lower dust temperatures (see Figure \ref{fig_beta_temperature_fig}) and, consequently, our sources would still satisfy the low--redshift SMG selection criterion.  It is worth noting that, when $\beta$ is left as a free parameters, most low--redshift SMGs have $\beta \sim 1.5$ (in agreement with \cite{Magnelli2012A&A...539A.155M} when they leave $\beta$ as a free parameter).  This also happens to many {\it H}--ATLAS sources, explaining the accumulation of red points in Figure \ref{fig_beta_temperature_fig} in the one--to--one relation (note that $\beta = 1.5$ is the minimum value allowed for $\beta$ in our fits).

\begin{figure}
\centering
\includegraphics[width=0.45\textwidth]{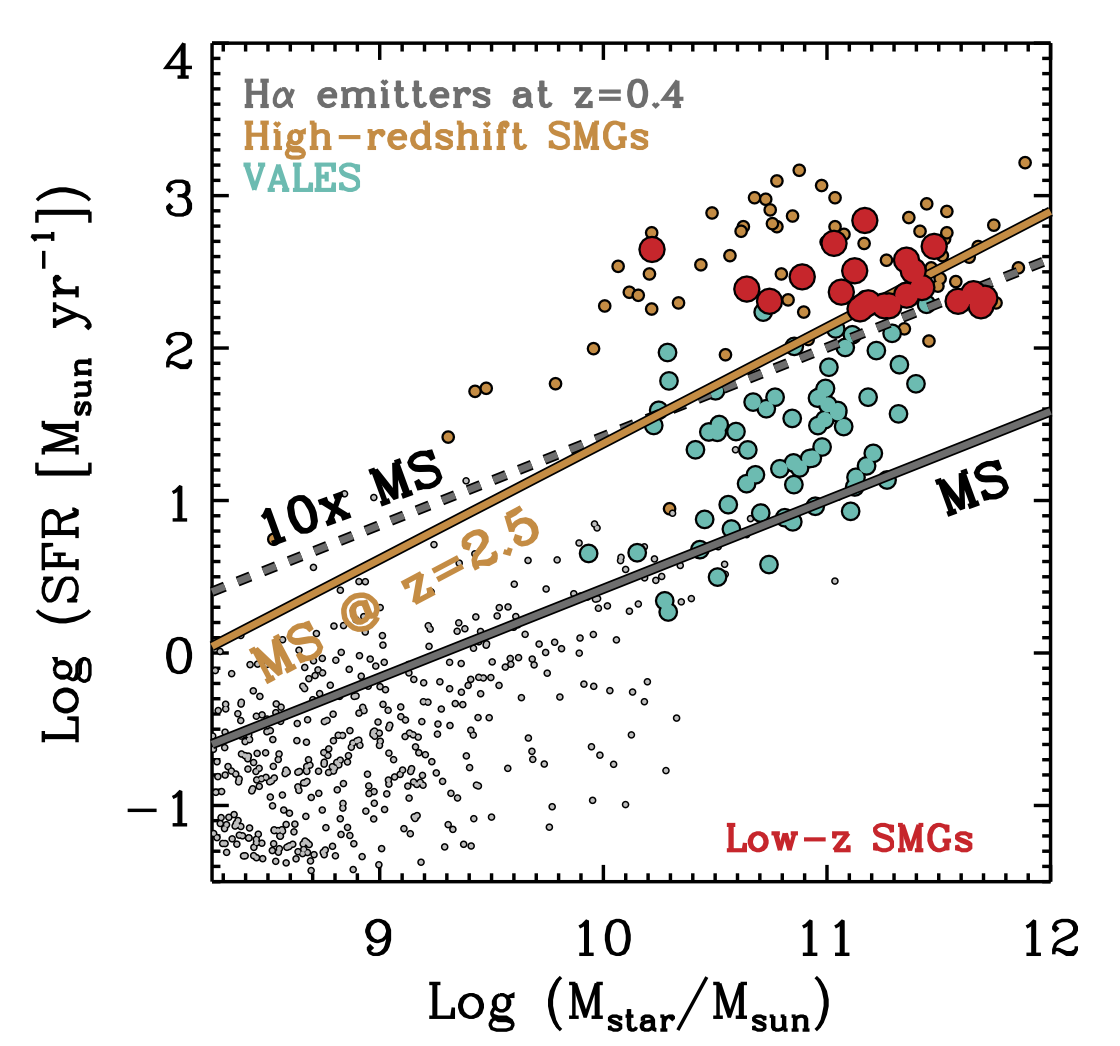} 
\caption{Location of our low--redshift SMGs in the star--formation rate versus stellar mass diagram, in comparison with other population of galaxies at comparable redshifts, including IR--bright sources from VALES \citep{Villanueva2017arXiv170509826V} and narrow--band selected H$\alpha$ emitters at $z \sim 0.4$ from \cite{Sobral2013MNRAS.428.1128S,Sobral2014MNRAS.437.3516S}.   For a reference, we show the location of high--redshift SMGs \citep{Rowlands2014MNRAS.441.1017R,daCunha2015ApJ...806..110D}.  We show with a solid line the star--formation main sequence (MS) at $z \sim 0.4$ from \cite{Speagle2014ApJS..214...15S}, and $10\times$ times the main sequence.  We also show the MS at $z \sim 2.5$.  We see that, as it happens to high--redshift, IR--luminous starbursts \citep{Rodighiero2011ApJ...739L..40R,daCunha2015ApJ...806..110D}, our low--redshift SMGs are well above the main sequence.  Actually,  our low--redshift SMGs are even further off the MS than high--redshift SMGs (note that the MS evolves with redshift, and for a given stellar mass, the MS at higher redshifts corresponds to higher SFRs).
              }
\label{sfr_mass_fig}
\end{figure}

\begin{figure*}
\centering
\includegraphics[width=0.90\textwidth]{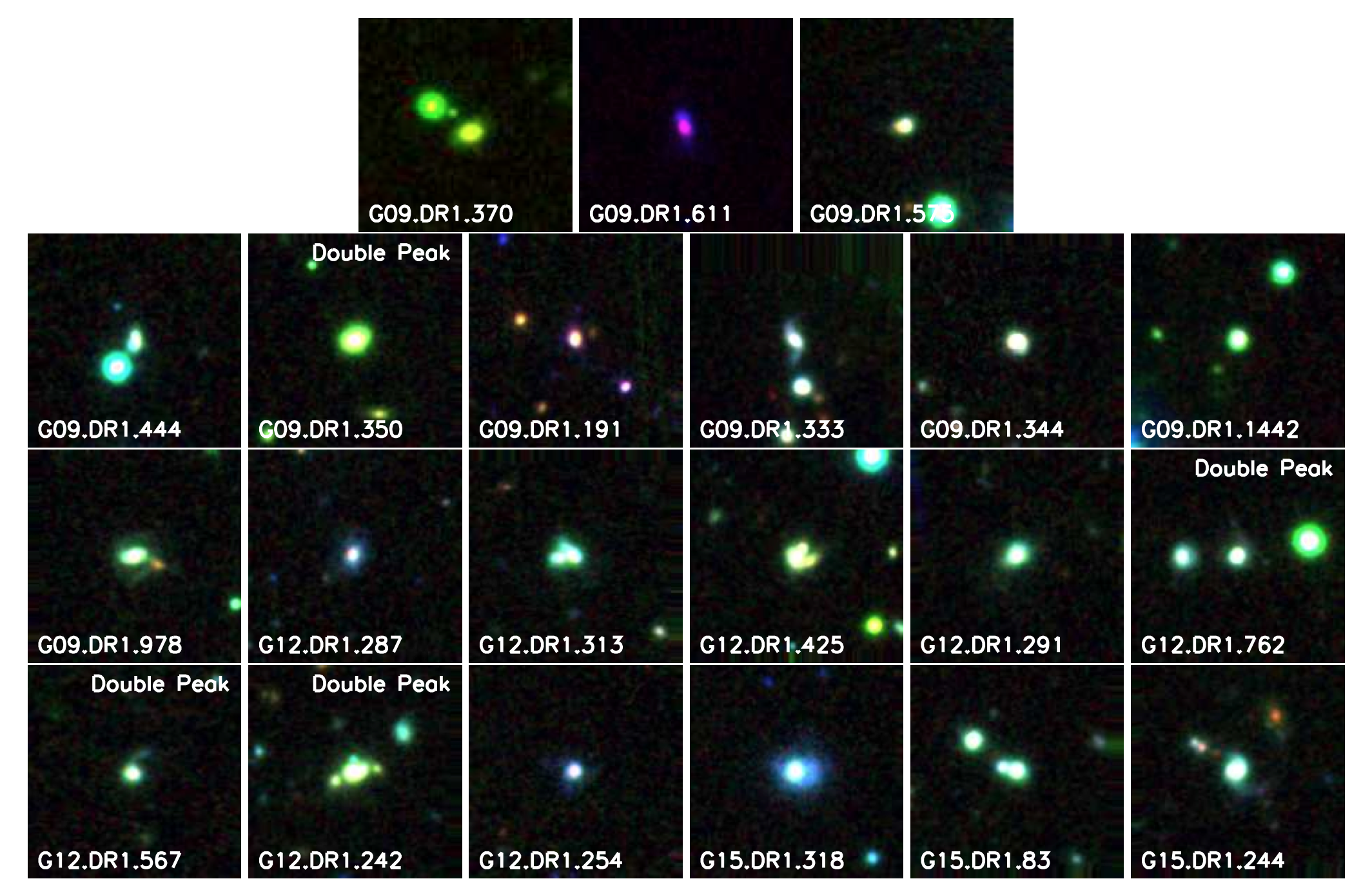} 
\caption{False--color images of all our low--redshift SMGs, created by using $r$, $z$, and $K_s$ imaging from the KIDS and VIKING surveys \citep{deJong2015A&A...582A..62D,Edge2013Msngr.154...32E}. Panels are $30''$ on each side, and are centered on the {\it Herschel} detections.  The size of the panels is comparable with the IRAM--30m half--power beam at $3 \, {\rm mm}$, and therefore there could contribution to the $^{12}$CO(1--0) emission from all the galaxies in these stamps, if they are at the same redshift as the low--redshift SMG in the center of the map.  We see a variety of morphologies, from relatively isolated sources (like G09.DR1.611, G12.DR1.287, or G12.DR1.254) to likely interacting systems (like G09.DR1.978, G12.DR1.242 or G15.DR1.83), although the majority of sources belong to the second group.  We also see a variety of colors, from very red (G09.DR1.611) to very blue sources (G15.DR1.318).  We indicate those source whose CO lines show a double--peak profile.  We note that the remarkable color of G09.DR1.611 is due to a slightly different morphology between the $r$ and $K_s$ imaging, with the $r$ imaging revealing that the optical emission is more extended than the near--IR one.
              }
\label{figure_optical_imaging_morphology}
\end{figure*}

\begin{figure*}[!t]
\centering
\includegraphics[width=0.90\textwidth]{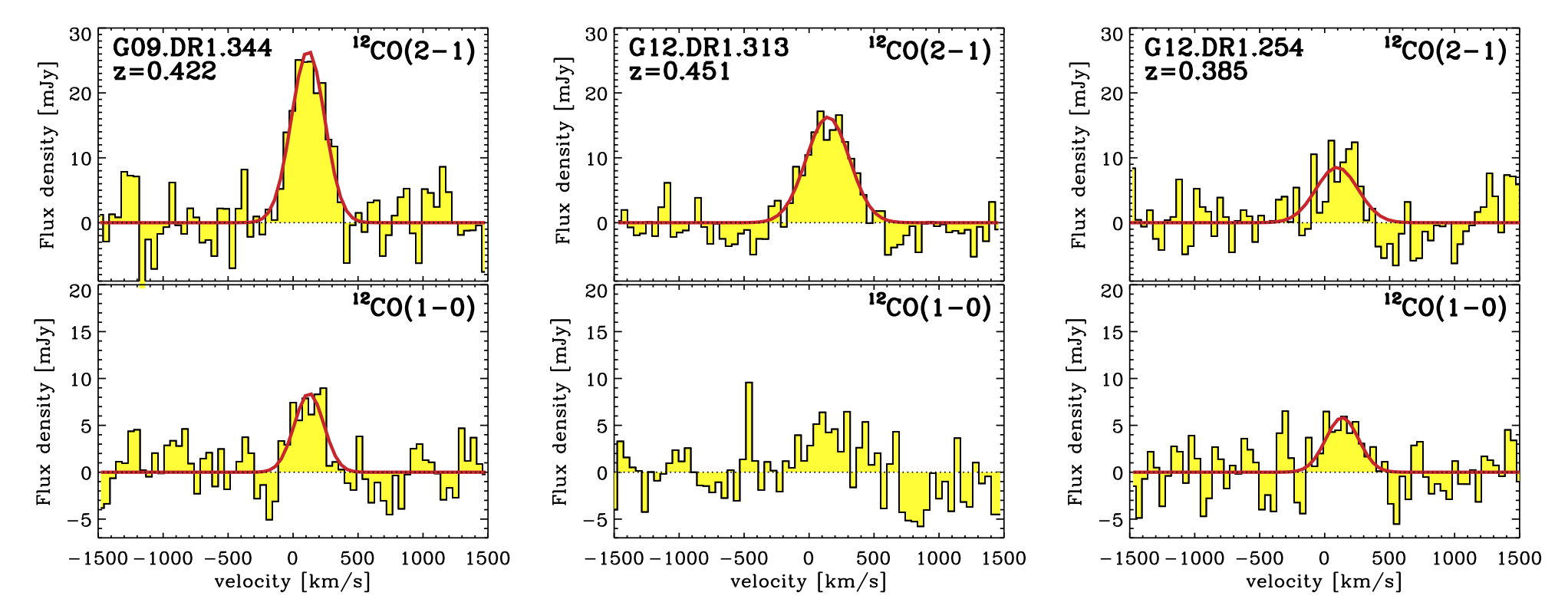}\\

\caption{Examples of the CO detection in our sample of low--redshift SMGs. We include in this figure three cases that illustrate the diversity of line brightnesses and ratios that we have found in our sample: from sources with bright CO emission and CO line ratio similar to high--redshift SMGs (G09.DR1.344 -- {\it left}), highly--excited sources (G12.DR10.313 -- {\it middle}) and sources with faint CO lines and low CO line ratio (G12.DR1.254 -- {\it right}). The spectra of all low--redshift SMGs observed with the IRAM--30m telescope are shown in the Appendix.
              }
\label{CO_specs_fig}
\end{figure*}

\section{Observations: IRAM--30m}\label{section_IRAM_observations}

The observations presented in this paper were carried out with the IRAM--30m telescope sited in Sierra Nevada, Spain, during April 2016, February 2017 and April 2017 in good to excellent weather conditions. This does not mean that the weather during the entire three observing runs was good, but instead that we only use the data taken in the 20\% of the time when the weather was good. The recent upgrade of the EMIR receivers allowed us to observe simultaneously the $^{12}$CO(2--1) and $^{12}$CO(1--0) transitions in all our sources. The observations were carried out in wobbler switching mode, with reference positions offset by $2'$ in azimuth. As backends, we used both FTS and WILMA. In this paper we will present the FTS data for all galaxies, except one of them (G12.DR1.762) for which the $^{12}$CO(2--1) emission detected with the FTS was affected by platforming so the WILMA spectrum is presented instead.  In all cases where a given transitions was covered by both WILMA and FTS we have checked that the line flux are in agreement within the uncertainties.

Typically, each source was observed for about $2\,{\rm h}$, although sources with brighter CO lines were observed for a shorter period.  Data reduction was carried out while observing, so whenever both $^{12}$CO(2--1) and $^{12}$CO(1--0) were clearly detected in a given source, we started observing another target. Since $^{12}$CO(1--0) is fainter than $^{12}$CO(2--1), the detection of $^{12}$CO(1--0) normally took longer than the detection of $^{12}$CO(2--1). Actually, as it will be discussed in \S \ref{section_CO_detections}, the $^{12}$CO(2--1) emission was detected in all the galaxies we have observed except in one (in which $^{12}$CO(1--0) was detected), whereas $^{12}$CO(1--0) was not detected in several sources due to the combination of a lack of sensitivity and high excitation of the molecular gas. The telescope half power beam at $3 \, {\rm mm}$ and $2 \, {\rm mm}$ is about $27''$ and $17''$, respectively. Since this is much larger than the optical sizes of our galaxies we do not expect to miss significant flux even in the $^{12}$CO(2--1) transition. However, this would not be true if some of our low--redshift SMGs would be formed by several physically--related components separated more than $\sim 15''$ ($\sim 80 \, {\rm kpc}$).  This effect could actually explain the relatively low CO line ratios that we find in some of our sources (see \S \ref{CO_SLED_section} and Figure \ref{CO_SLED_fig}), but interferometric observations would be needed to confirm this.

\section{Results}\label{results_of_the_paper}

\subsubsection{Stellar mass and SFR--mass relation}

We have estimated the stellar masses of our low--redshift SMGs by fitting their observed multi--wavelength photometry with MAGPHYS \citep{daCunha2008MNRAS.388.1595D} with the extended high--redshift priors \citep{daCunha2010A&A...523A..78D,Rowlands2014MNRAS.441.1017R,daCunha2015ApJ...806..110D}.  The photometry includes GALEX, SDSS, VISTA, WISE and {\it Herschel} data. The derived stellar masses are shown in Tables \ref{table_full_sample} and \ref{table_flux_density_and_redshift}, and range from $\sim 2 \times 10^{10}$ to $\sim 5 \times 10^{11}\,M_\odot$. The stellar masses of our low--redshift SMGs are compatible to those found in high--redshift SMGs (see Figure \ref{sfr_mass_fig}). We note that the determination of the stellar mass in our galaxies is affected by the number of uncertainties concerning the stellar mass determination in such sources due to the complexity of the star formation histories and dust obscuration, as it also happens in high--redshift SMGs \citep{Engel2010ApJ...724..233E,Michalowski2010A&A...514A..67M,Hainline2011ApJ...740...96H}. 


Figure \ref{sfr_mass_fig} shows the location of low--redshift SMGs in the classical SFR versus stellar mass diagram, where we also show the main sequence (MS) of galaxies at $z \sim 0.4$ and $z \sim 2.5$ \citep{Speagle2014ApJS..214...15S}.  Sources in the VALES sample, H$\alpha$ emitters at $z \sim 0.4$ from \cite{Sobral2013MNRAS.428.1128S,Sobral2014MNRAS.437.3516S} and high--redshift SMGs \citep{Michalowski2017MNRAS.469..492M} are also included for a reference.  It can be seen that, as it happens to high--redshift starbursts \citep{Rodighiero2011ApJ...739L..40R}, low--redshift SMGs are clear interlopers to the MS. Actually, low--redshift SMGs are located further off the MS than their high--redshift counterparts.  This is because the stellar mass and SFR of low-- and high--redshift SMGs are similar, but the MS evolves with redshift in the sense that for a given stellar mass, the MS at higher redshift is associated to higher SFR \citep{Elbaz2011A&A...533A.119E}.

\subsection{Optical/near--IR morphology}\label{section_optical_moph}

\begin{table*}[!t]
\caption{\label{CO_lines_properties_TABLE}CO line properties in our low--redshift SMGs}
\centering
\begin{tabular}{cccccccccccccccccccc}
\hline
Source 	& $I_{\rm CO(2-1)}$ 			& ${\rm FWHM_{CO(2-1)}}$		& $I_{\rm CO(1-0)}$ & ${\rm FWHM_{CO(1-0)}}$	\\
  		& $[{\rm Jy \, km \, s^{-1}}]$ 	& $[{\rm km \, s^{-1}}]$	& $[{\rm Jy \, km \, s^{-1}}]$ & $[{\rm km \, s^{-1}}]$	\\
\hline\hline
	G09.DR1.370	&	$5.4 \pm 1.8$	&	$560 \pm 141$	& 	$< 1.8$			& 	--				\\
	G12.DR1.254	&      $3.5 \pm 1.2$	&	$381 \pm 98$	&	$1.8 \pm 0.6$	&     	$299 \pm 83$ 		\\
	G09.DR1.350	&	$7.2 \pm 0.9$ 	&	$425 \pm 38$	&	$2.8 \pm 0.7$	&	$389 \pm 67$		\\
	G09.DR1.344	&     	$8.5 \pm 1.1$	&	$302 \pm 30$	&	$2.4 \pm 0.7$	&	$267 \pm 58$		\\
	G15.DR1.244	&	$4.2 \pm 0.8$	&	$302 \pm 45$	&	$1.7 \pm 0.8$	&	$172 \pm 63$		\\
	G12.DR1.291	&	$6.0 \pm 1.2$	&	$524 \pm 77$	&	$< 1.4$			&	--				\\
	G12.DR1.313	&	$6.8 \pm 0.7$	&	$392 \pm 32$	&	$< 1.3$			&	--				\\
	G12.DR1.762	&	$4.4 \pm 1.1$	&	$506 \pm 91$	&	$3.0 \pm 0.9$	&	$369 \pm 79$		\\	
	G15.DR1.83	&	$6.3 \pm 1.1$	&	$298 \pm 41$	&	$< 1.3$			&	--				\\
	G09.DR1.575	&	$3.5 \pm 0.4$	&	$259 \pm 24$	&	$< 1.0$			&	--				\\	
	G12.DR1.567	&	$6.0 \pm 1.0$	&	$365 \pm 45$	&	$< 1.4$			&	--				\\
	G09.DR1.333	&	$10.8 \pm 1.8$ &	$370 \pm 44$	&	$< 2.0$			&	--				\\
	G09.DR1.191	&	$< 5.8$			&	--			&	$3.9 \pm 1.2$	&	$365 \pm 85$		\\
	G12.DR1.287	&	$2.5 \pm 0.7$	&	$242 \pm 50$	&	$2.3 \pm 1.0$	&	$340 \pm 111$		\\
	G12.DR1.425	&	$5.0 \pm 1.0$	&	$377 \pm 56$	&	$< 1.2$			&	--				\\
	G12.DR1.242	&	$2.3 \pm 0.7$	&	$384 \pm 82$	&	$3.1 \pm 0.9$	&	$561 \pm 117$		\\		
\hline
\hline
\end{tabular}
\end{table*}

We show in Figure \ref{figure_optical_imaging_morphology} the false--color images of our low--redshift SMGs, built from the available KIDS and VIKING near--IR imaging in the GAMA fields \citep{deJong2015A&A...582A..62D,Edge2013Msngr.154...32E}.  It can be seen that about 60\% of our low--redshift SMGs are likely formed by at least two sources in interaction or belonging to groups of galaxies (like G09.DR1.978, G12.DR1.313, or G12.DR1.242), and the remaining 40\% are relatively isolated sources (see for example G09.DR1.344, G12.DR1.287 or G15.DR1.318).  We also see a noticeable range of colors, from very red (G09.DR1.611) to very blue sources (G12.DR1.287 or G15.DR1.318).

The diversity of colors and morphologies that we see in our low--redshift SMG sample is in agreement with those works reporting that most SMGs are mergers \citep{Tacconi2008ApJ...680..246T,Engel2010ApJ...724..233E}, but also with those suggesting that major mergers are not the dominant driver of the high--redshift SMG population \citep{Narayanan2015Natur.525..496N,Hodge2015ApJ...798L..18H,Michalowski2017MNRAS.469..492M}.  Actually, our results might indicate, insomuch as our low-redshift sample can be taken as indicative of high-redshift SMGs, that high--redshift SMGs might be triggered by more than one mechanism.  


It is important to point out that the fact that some of our low--redshift SMGs seem to be disk--like, isolated sources and the fact that they are clear outliers of the MS (see Figure \ref{sfr_mass_fig}) is in opposition to the classical thought that the outliers of the MS are interacting sources. Our results seem to suggest that outliers of the MS can also be disk--like galaxies, at least at low redshift.

\begin{figure*}
\centering
\includegraphics[width=0.90\textwidth]{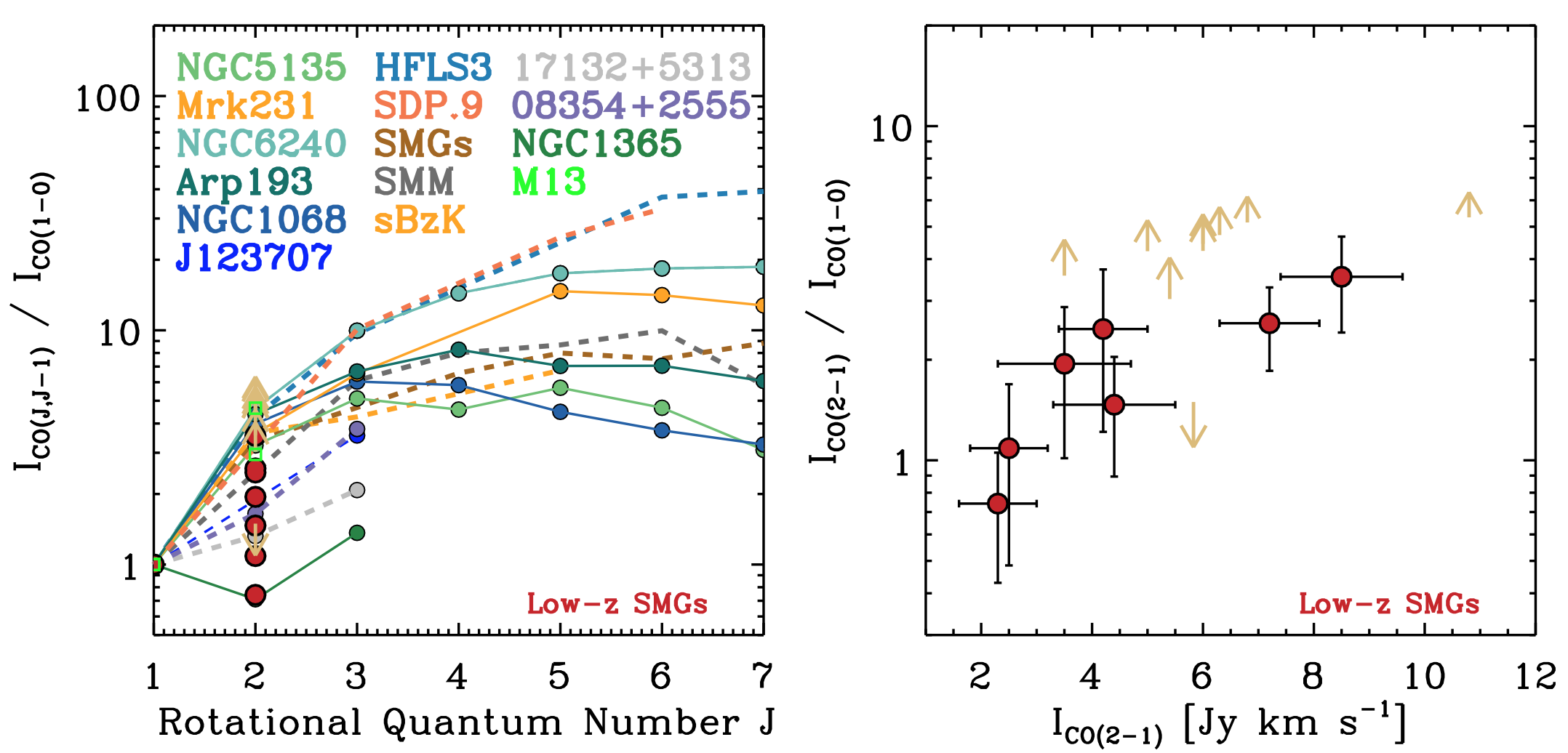} 
\caption{{\it Left}: CO line ratio of our low--redshift SMGs, compared to those for other populations of galaxies at different redshifts \citep{Lupu2012ApJ...757..135L,Vlahakis2015ApJ...808L...4A,Harris2012ApJ...752..152H,Frayer2011ApJ...726L..22F,Omont2013A&A...551A.115O,Oteo2017arXiv170105901O,Bothwell2013MNRAS.429.3047B,Danielson2011MNRAS.410.1687D,Riechers2013Natur.496..329R,Meijerink2013ApJ...762L..16M,Papadopoulos2014ApJ...788..153P,Rosenberg2015ApJ...801...72R,Papadopoulos2012MNRAS.426.2601P}. Red dots represent low--redshift SMGs whose $^{12}$CO(2--1) and $^{12}$CO(1--0) transitions have been detected, while arrows indicate upper of lower limits for sources where one of the lines has not been detected.  We also represent (green squares) the two ULIRGs in \cite{Magdis2013A&A...558A.136M} which would have been selected as low--redshift SMGs and have $^{12}$CO(2--1) and $^{12}$CO(1--0) detections. We see that there is a wide range of CO excitations in our sample, from low--excited sources such as G09.DR1.191 or G12.DR1.762 to highly--excited sources such as G12.DR10.313 or G15.DR1.83 (see spectra in Figures \ref{CO_specs_fig_1} and \ref{CO_specs_fig_2}). This suggests that assuming the average line excitation for SMGs (or any other fixed CO line ratio) to measure the total molecular gas of our sources could lead to significant uncertainties. {\it Right}:  Line ratio between the $^{12}$CO(2--1) and $^{12}$CO(1--0) transitions as a function of the $^{12}$CO(2--1) line flux for our low--redshift SMGs.  Upper and lower limits correspond to cases where one of the two transitions has not been detected.  This panel is shown to highlight the wide range of CO excitation in our low--redshift SMGs, now including the error bars which were not shown in the left panel for the sake of clarity.  In both panels, line flux ratios are shown in the $y$ axes.
              }
\label{CO_SLED_fig}
\end{figure*}

\subsection{CO detections}\label{section_CO_detections}

We have detected at least one CO line in all 16 sources that we have observed with the IRAM--30m telescope (line properties shown in Table \ref{CO_lines_properties_TABLE}). Figure \ref{CO_specs_fig} shows three cases of CO detections: one source with clear detections of both $^{12}$CO(2--1) and $^{12}$CO(1--0), an excited source with bright $^{12}$CO(2--1) emission and no $^{12}$CO(1--0) detection, and a source with faint $^{12}$CO(2--1) and $^{12}$CO(1--0) detections. The fluxes of the detected lines have been derived from Gaussian fits to the line profile. Whenever a line has not been detected, we have derived the $3 \sigma$ upper limit on the line flux assuming that the width of the $^{12}$CO(2--1) and $^{12}$CO(1--0) emission are the same. The lines fluxes have been converted to line luminosities ($L'_{\rm CO}$) following \cite{Carilli2013ARA&A..51..105C}. These are shown in Table \ref{table_flux_density_and_redshift}.

We have estimated the molecular gas masses for our sample from the $^{12}$CO(1--0) luminosity ($L'_{\rm CO(1-0)}$) wherever it has been detected. For galaxies without $^{12}$CO(1--0) detection we have measured the molecular gas mass by estimating the $^{12}$CO(1--0) luminosity assuming the average line luminosity ratio for SMGs \citep{Carilli2013ARA&A..51..105C}. The molecular gas masses (see values in Table \ref{table_flux_density_and_redshift}) have been derived from the CO luminosities by using the $\alpha_{\rm CO}$ conversion factor for local ULIRGs, $\alpha_{\rm CO} = 0.8 M_\odot / \, ({\rm K \, km \, s^{-1} \, pc^2})$ \citep{Downes1998ApJ...507..615D}.  We need to point out here that, as it will be discussed in \S \ref{CO_SLED_section}, there is a significant range of CO excitation in our sample of low--redshift SMGs, meaning that the distribution of the $^{12}$CO(2--1)/$^{12}$CO(1--0) line ratios is very wide and making the determination of the molecular gas mass from $^{12}$CO(2--1) uncertain (not only here, but also in all previous work).  We see that our low--redshift SMGs have massive molecular gas reservoirs, with masses ranging from $M_{\rm H_2} \sim 9 \times 10^9 \, M_\odot$ to $M_{\rm H_2} \sim 2.5 \times 10^{10} \, M_\odot$.  The most massive molecular gas reservoirs are as massive as those found in high--redshift SMGs with $^{12}$CO(1--0) detections \citep[see for example][]{Ivison2011MNRAS.412.1913I,Riechers2011ApJ...739L..31R,Walter2011ApJ...730...18W}.  When comparing the molecular gas masses with high--redshift SMGs we should keep in mind that $^{12}$CO(1--0) has only been detected in the most massive and extreme high--redshift SMGs, whose total IR luminosities overlap only slightly with the $L_{\rm IR}$ in our sample.  It is plausible that the significant sample of high--redshift SMGs with $L_{\rm IR} \sim 10^{12} \, L_\odot$ which have not been detected in $^{12}$CO(1--0) has molecular gas masses similar to our low--redshift SMGs if the $L_{\rm IR} - L_{\rm CO}$ relation holds at all redshifts.

In addition to the wide range of line ratios, we also see a variety of line profiles.  Four of our 16 low--redshift SMGs show clear double--peak profiles.  The percentage of double--peak CO profiles is compatible to that found in a sample of 32 high--redshift SMGs with CO detections \cite{Bothwell2013MNRAS.429.3047B}, and lower than those reported, for example, in \cite{Greve2005MNRAS.359.1165G,Tacconi2006ApJ...640..228T,Tacconi2008ApJ...680..246T} or \cite{Engel2010ApJ...724..233E} in smaller samples.  The presence of double--peaked CO emission might be an indication of two kinematically distinct components, although it could also be due to a rotating disk.  In Figure \ref{figure_optical_imaging_morphology} we have indicated the four galaxies in our sample with double--peak CO profiles.  We see that the near--IR images of two of these sources (G12.DR1.762 and G12.DR1.242) show evidence of interactions, whereas the other two (G09.DR1.350 and G12.DR1.567) seem to be isolated disks.  Therefore, our results suggest that double--peaked CO line profiles are not always an indication of two interacting components, and that caution should be taken when interpreting the morphology of high--redshift SMGs by using the CO line profiles.

\subsection{The CO line ratios}\label{CO_SLED_section}

We study in this section the CO line ratios of our 16 low--redshift SMGs with CO detections, which is shown in Figure \ref{CO_SLED_fig}. The lower limits on the CO line ratio represent sources where $^{12}$CO(1--0) has not been detected, whereas the upper limit corresponds to the source whose $^{12}$CO(2--1) has not been detected (G09.DR1.191 -- see Figure \ref{CO_specs_fig_2}). We compare the line ratio of our low--redshift SMGs with those for different populations of galaxies, including lensed ULIRGs at high redshift \citep{Lupu2012ApJ...757..135L,Vlahakis2015ApJ...808L...4A,Harris2012ApJ...752..152H,Frayer2011ApJ...726L..22F,Omont2013A&A...551A.115O,Oteo2017arXiv170105901O}, the classical population of high--redshift SMGs \citep{Bothwell2013MNRAS.429.3047B}, the lensed SMG SMM J2135$-$0102 at $z \sim 2.3$ \citep[labelled as SMM,][]{Danielson2011MNRAS.410.1687D}, HFLS3 at $z \sim 6.34$ \citep{Riechers2013Natur.496..329R} and several local (U)LIRGs \citep{Meijerink2013ApJ...762L..16M,Papadopoulos2014ApJ...788..153P,Rosenberg2015ApJ...801...72R,Papadopoulos2012MNRAS.426.2601P}.  Note that since we only have observed two $^{12}$CO lines we do not attempt to model the CO line ratio of our galaxies, but instead we only discuss the observed line ratios and their implications on the nature of our sources and the determination of the molecular gas mass.  We note that caution must be taken when comparing CO line ratios without the use of modeling, as high--redshift SMGs see a hotter CMB that might alter the observed CO line ratios, and this effect can not be corrected in a simple way due to its non-linearity \citep{Zhang2016RSOS....360025Z}.

 \begin{table*}
\caption{\label{table_flux_density_and_redshift}Properties of our sample of low-$z$ SMGs with CO detections}
\centering
\begin{tabular}{cccccccccccccccccccc}
\hline
Source 	&  $L'_{\rm CO(2-1)}$	& $L'_{\rm CO(1-0)}$	&	$M_{\rm H2}$\tablenotemark{(a)} &	$M_{\rm star}$ &	$M_{\rm dust}$	& ${\rm SFR}$	\\
  		& $[{\times 10^{10}\, \rm K \, km \, s^{-1} \, {\rm pc}^2}]$	 	& $[\times 10^{10} {\rm K \, km \, s^{-1} \, {\rm pc}^2}]$	&	[$\times 10^{10} \, M_\odot$] & [$ \times 10^{11}\,M_\odot$] & [$ \times 10^{8}\,M_\odot$] & $[M_\odot \, {\rm yr}^{-1}]$ \\
\hline\hline
	G09.DR1.370			&	$1.0 \pm 0.3$ 		& 	$< 1.5$		&	$\sim 1.0$ 		& $\sim 3.9 $&	$\sim 2.8$			&	$\sim 210$	\\
	G12.DR1.254			&  	$0.7 \pm 0.2$		&   	$1.4 \pm 0.5$	&	$\sim 1.1$ 		& $\sim 0.2 $&	$\sim 3.7$			&	$\sim 450$	\\
	G09.DR1.350			&	$1.4 \pm 0.2$		&	$2.2 \pm 0.5$	&	$\sim 1.8$ 		& $\sim 3.0 $&	$\sim 4.8$			&	$\sim 470$	\\
	G09.DR1.344			&	$2.0 \pm 0.3$		&	$2.2 \pm 0.6$	&	$\sim 1.8$ 		& $\sim 2.3 $&	$\sim 2.5$			&	$\sim 230$	\\
	G15.DR1.244			&	$1.0 \pm 0.2$		&	$1.7 \pm 0.7$	&	$\sim 1.4$ 		& $\sim 4.4 $&	$\sim 7.7$			&	$\sim 230$	\\
	G12.DR1.291			&	$1.6 \pm 0.3$		&	$< 1.1$		&	$\sim 1.5$ 		& $\sim 2.6 $&	$\sim 5.9$			&	$\sim 250$	\\
	G12.DR1.313			&	$1.7 \pm 0.2$		&	$< 1.1$		&	$\sim 1.6$ 		& $\sim 3.6 $&	$\sim 7.2$			&	$\sim 190$	\\
	G12.DR1.762			&	$1.2 \pm 0.3$		&	$3.2 \pm 0.9$	&	$\sim 2.5$ 		& $\sim 1.5 $&	$\sim 3.6$			& 	$\sim 200$	\\	
	G15.DR1.83			&	$1.7 \pm 0.3$		&	$< 1.1$		&	$\sim 1.6$ 		& $\sim 1.5 $&	$\sim 7.0$			&	$\sim 700$	\\
	G09.DR1.575			&	$1.0 \pm 0.1$		&	$< 0.8$		&	$\sim 0.9$ 		& $\sim 0.8 $&	$\sim 5.0$			&	$\sim 300$	\\	
	G12.DR1.567			&	$1.8	\pm 0.3$		&	$< 1.2$		&	$\sim 1.7$ 		& $\sim 1.1 $&	$\sim 5.9$			&	$\sim 240$	\\
	G09.DR1.333			&	$2.4\pm  0.4$		&	$< 1.7$		&	$\sim 2.3$			& $\sim 0.5 $&	$\sim 5.9$			&	$\sim 210$	\\
	G09.DR1.191			&	--				&	$2.8 \pm 1.0$	&	$\sim 2.2$			& $\sim 1.4 $&	$\sim 4.8$			&	$\sim 190$	\\
	G12.DR1.287			&	$0.5 \pm 0.1$		&	$1.9 \pm 0.8$	&	$\sim 1.6$			& $\sim 1.4 $&	$\sim 5.4$			&	$\sim 180$	\\
	G12.DR1.425			&	$1.1 \pm 0.2$		&	$< 1.0$		&	$\sim 1.0$			& $\sim 4.9 $&	$\sim 6.5$			&	$\sim 190$	\\
	G12.DR1.242			&	$0.5\pm 0.1$		&	$2.8 \pm 0.8$	&	$\sim 2.2$			& $\sim 5.1 $&	$\sim 8.0$			&	$\sim 220$	\\		
\hline
\hline
\tablenotetext{1}{The gas masses have been obtained by assuming the CO conversion factor traditionally used for local ULIRGs and high--redshift dusty starbursts, $\alpha_{\rm CO} = 0.8 M_\odot / \, ({\rm K \, km \, s^{-1} \, pc^2})$ \citep{Downes1998ApJ...507..615D}.}
\end{tabular}
\end{table*}

It can be clearly seen that there is a wide range of CO excitation in our sample, from galaxies with highly--excited molecular gas revealed by their high $^{12}$CO(2--1)/$^{12}$CO(1--0) line ratios (for example G12.DR10.313 or G15.DR1.83) to others with much less excited molecular gas and very low CO line ratios (see for example G09.DR1.191 where $^{12}$CO(2--1) has not been detected).  The wide range of CO excitation found in our low--redshift SMGs is compatible with that for other IR--bright galaxies, both at low and high redshift. Our low--redshift SMGs can be as excited as the most excited sources in the literature such as HFLS\,3 at $z \sim 6.3$, or can have low CO ratios compatible with the least excited local sources such as 17132+5313 \citep{Papadopoulos2012MNRAS.426.2601P}.  The two ULIRGs in \cite{Magdis2013A&A...558A.136M} that would be selected as low--redshift SMGs and with $^{12}$CO(2--1) and $^{12}$CO(1--0) are as excited as the most excited low--redshift SMGs, although we note that one of the sources in \cite{Magdis2013A&A...558A.136M} has a $^{12}$CO(1--0) emission much narrower than the $^{12}$CO(2--1), which could mean that the line ratio is over--estimated.

The right panel of Figure \ref{CO_SLED_fig} represents the $^{12}$CO(2--1)/$^{12}$CO(1--0) line flux ratio of our low--redshift SMGs as a function of their $^{12}$CO(2--1) line flux.  This panel is included to clarify and support the wide range of line excitation in our low--redshift SMGs, which is considerably higher than the uncertainties and further reinforced with the lower and upper limits.  There seems to be a trend of higher excitation with increasing $^{12}$CO(2--1) line flux.  However, this trend might be due to the limited range of total IR luminosities of the galaxies in our sample, producing that sources with brighter $^{12}$CO(2--1) lines are more excited (assuming that the $L_{\rm IR} - L'_{\rm CO(1-0)}$ holds for our low--redshift SMGs -- see \S \ref{lir_co_correlation}).

We should note that one of the problems in the study of the CO line ratios is the different size of the IRAM--30m beam in our $^{12}$CO(2--1) and $^{12}$CO(1--0) observations.  As we point out in \S \ref{section_optical_moph}, all galaxies on each stamp (if they are at the same redshift as the low--redshift SMG in the center) could be contributing to the $^{12}$CO(1--0) emission.  However, since the beam is smaller in the $^{12}$CO(2--1) observations, it could happen that we might lose some $^{12}$CO(2--1) line flux from sources (again, if they are at the same redshift as the central low--redshift SMG) close to the edge of the $^{12}$CO(1--0) beam, lowering the CO line ratio.  This might explain the low CO ratio seen in some of our galaxies, like G12.DR1.242 that might have a companion located about $\sim10''$ away from the low--redshift SMGs observed with IRAM--30m.  However, there are sources like G12.DR1.287, which have a relatively low CO ratio and are relatively isolated.

Our results reinforce the idea, put forward many times, that extrapolations from mid--$J$ $^{12}$CO to $^{12}$CO(1--0) could be highly uncertain.  Actually, even extrapolating from $^{12}$CO(2--1) to $^{12}$CO(1--0) for galaxies without $^{12}$CO(1--0) detections could be highly uncertain, up to a factor of $\sim 10 \times$.  Such extrapolation would be required to, for example, measure the molecular gas mass from $^{12}$CO(1--0) when this transition has not been detected or cannot be observed (if redshift prevents to do so).  The uncertainty in the CO line ratio adds to the uncertainty in the $\alpha_{\rm CO}$ factor to convert from the CO luminosity to the molecular gas mass, making determinations of the molecular gas mass (and related properties such as molecular gas surface densities, star--formation efficiency or gas depletion times) from mid--$J$ CO lines highly uncertain.

\subsection{The $L_{\rm IR} - L'_{\rm CO}$ correlation}\label{lir_co_correlation}

\begin{figure}
\centering
\includegraphics[width=0.45\textwidth]{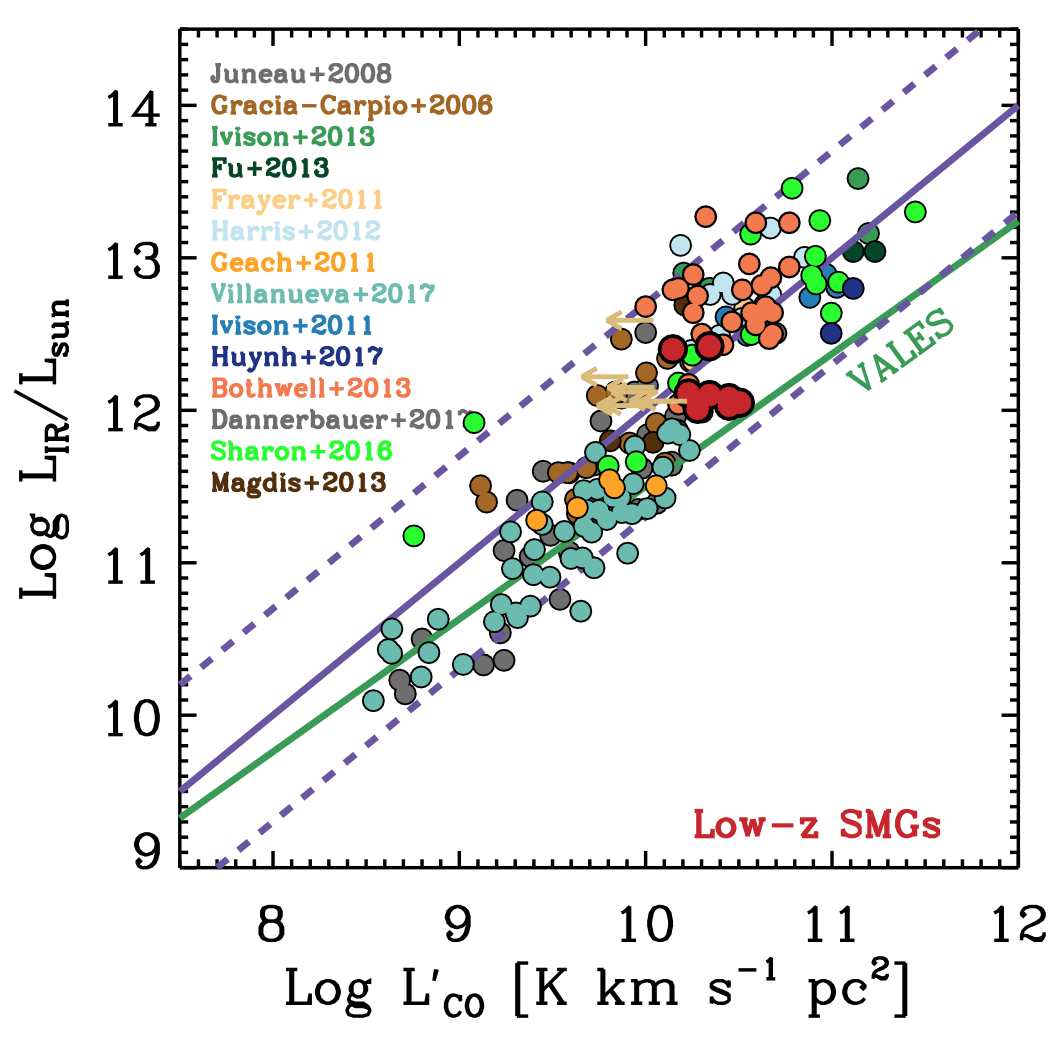} 
\caption{Total IR against $^{12}$CO(1--0) luminosity for our sample of low--redshift SMGs (both detections and upper limits, the latter indicated with the horizontal arrows), along with the values reported in the literature for other populations at different redshifts \citep{Juneau2009ApJ...707.1217J,GraciaCarpio2006ApJ...640L.135G,Magdis2013A&A...558A.136M,Ivison2013ApJ...772..137I,Fu2013Natur.498..338F,Frayer2011ApJ...726L..22F,Harris2012ApJ...752..152H,Geach2011ApJ...730L..19G,Villanueva2017arXiv170509826V,Ivison2011MNRAS.412.1913I,Bothwell2013MNRAS.429.3047B,Huynh2017MNRAS.467.1222H,Sharon2016ApJ...827...18S,Dannerbauer2017arXiv170105250D}.  We represent the $L_{\rm IR} - L'_{\rm CO}$ correlation reported in \cite{Greve2014ApJ...794..142G} with a purple solid line (purple dashed lines show a $\pm 0.7 \, {\rm dex}$ spread with respect to that relation) and \cite{Villanueva2017arXiv170509826V} with a green solid line.  We only represent here galaxies whose $^{12}$CO(1--0) emission has been detected, except \cite{Bothwell2013MNRAS.429.3047B}, so we minimize the effect of the uncertain line ratio between mid--$J$ lines and $^{12}$CO(1--0) lines. Actually, as it has been reported in Figure \ref{CO_SLED_fig}, the significant range of line excitations makes the extrapolation from $J > 1$ to $^{12}$CO(1--0) be uncertain.  Note that this diagram is equivalent to the relation between the SFR and the molecular gas mass.  We argue that most points in the literature can be fitted by a single linear relation with a significant scatter, unlike previous findings suggesting that {\it normal} star--forming galaxies and IR--bright starbursts follow different trends.  
              }
\label{lumi_lumi_SMGs}
\end{figure}

Using the $^{12}$CO(1--0) detections for our low--redshift SMGs we can study the relation between the total IR and CO luminosity, which has been analyzed in many previous work at different redshifts and luminosities ranges \citep{Greve2005MNRAS.359.1165G,Riechers2006ApJ...650..604R,Daddi2010ApJ...714L.118D,Genzel2010MNRAS.407.2091G,Ivison2011MNRAS.412.1913I,Greve2014ApJ...794..142G}.  We note that the $L_{\rm IR} - L'_{\rm CO}$ relation is equivalent to the relation between the SFR and the molecular gas mass (with the uncertainties resulting from the conversion from observables to physical quantities). The $L_{\rm IR} - L'_{\rm CO}$ relation for our low--redshift SMGs is shown in Figure \ref{lumi_lumi_SMGs}, where we also include other populations of galaxies at different redshifts.  Except for \cite{Bothwell2013MNRAS.429.3047B}, we plot literature data which are based on the detection of $^{12}$CO(1--0) detection. This minimizes the effect of the conversion from mid--$J$ to $J = 1-0$ CO luminosity, which has been actually shown in \S \ref{CO_SLED_section} to be highly uncertain even from the $J = 2-1$ to the $J = 1-0$ transitions.  We point out that the total IR luminosities in \cite{Bothwell2013MNRAS.429.3047B} have been measured from radio continuum emission, and not from FIR photometry as in our low--redshift SMGs.

\begin{figure}[!t]
\centering
\includegraphics[width=0.45\textwidth]{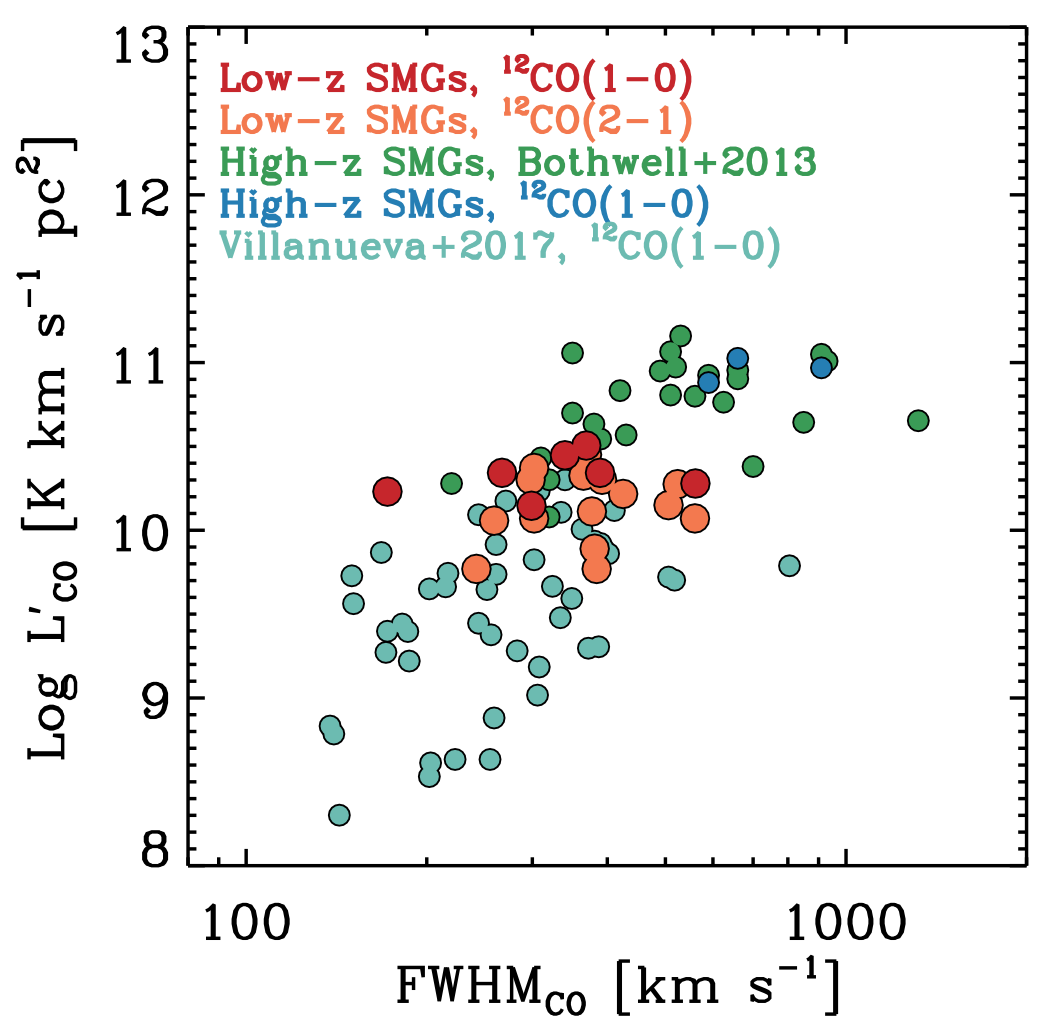} 
\caption{CO luminosity versus the CO line width (measured as its FWHM) for low--redshift SMGs, sources in the VALES survey \citep{Villanueva2017arXiv170509826V}, and high--redshift SMGs \citep{Ivison2011MNRAS.412.1913I,Bothwell2013MNRAS.429.3047B}.  We include in this plot measurements of different CO transitions, since most high--redshift SMGs have only been observed at $J > 3$.  We see that there is a relation between both parameters, in agreement with trends reported by other work \citep[see for example][]{Harris2012ApJ...752..152H}.  Such trend is not visible in our sample alone due to the narrow range in CO luminosities, which actually shows that the $L'_{\rm CO} - {\rm FWHM}$ relation is not tight.  We see that, despite the median FWHM for low--redshift SMGs is lower than for high--redshift SMGs (likely because only the most IR--luminous high--redshift SMGs have been followed-up in CO -- see Figure \ref{fig_CO_line_width}), there is a clear overlap between the two populations.
              }
\label{fig_CO_line_width}
\end{figure}

First, taking together all previous points reported in the literature, we see that despite there is a correlation between line and continuum luminosities, it is not tight, but instead has a significant spread. We include in Figure \ref{lumi_lumi_SMGs} two linear fits proposed in the literature, one by \cite{Greve2014ApJ...794..142G} and the other by \cite{Villanueva2017arXiv170509826V}. \cite{Greve2014ApJ...794..142G} obtained their correlation by using a sample of local ($z < 0.1$) (U)LIRGs and high--redshift SMGs with robust CO detections. \cite{Villanueva2017arXiv170509826V} found their correlation by using a sample of low--redshift ($z < 0.35$) IR--bright sources selected from {\it H}--ATLAS with ALMA $^{12}$CO(1--0) observations. The difference between these two recently proposed relations already highlights the spread of the $L_{\rm IR} - L'_{\rm CO}$ relation.

The location of our low--redshift SMGs with $^{12}$CO(1--0) detections in the $L_{\rm IR} - L'_{\rm CO}$ diagram is compatible with the spread of that relation and, therefore, with the location of the similarly IR--bright galaxies at both high and low redshift. We also represent in Figure \ref{lumi_lumi_SMGs} the upper limits corresponding to the low--redshift SMGs without $^{12}$CO(1--0) detections. Again, these upper limits suggest that the $L_{\rm IR} - L'_{\rm CO}$ relation has a significant spread. For a fixed total IR luminosity, the luminosity of the $^{12}$CO(1--0) transition can vary as much as one order of magnitude.

Figure \ref{lumi_lumi_SMGs} also shows that CO detections in high--redshift SMGs are limited to the most luminous sources. This is likely the reason why our low--redshift SMGs have lower CO luminosities than high--redshift SMGs.  In this way, in order to make the fairest comparison between low--redshift and high--redshift SMGs we would need both more low--redshift SMGs at the luminous end and more high--redshift SMGs with $^{12}$CO(1--0) detections which are less luminous than those studied so far.

We show in Figure \ref{fig_CO_line_width} the relation between the CO luminosity and linewidth for our low--redshift SMGs in comparison to those for high--redshift SMGs and several low--redshift galaxies.  In this figure we include galaxies with both $^{12}$CO(1--0), $^{12}$CO(2--1) and higher--$J$ CO transitions.  We see that there is a correlation between both parameters (with a quite significant spread of up to one order of magnitude), as it has been reported in previous work \citep[see for example][]{Harris2012ApJ...752..152H,Bothwell2013MNRAS.429.3047B}, which extended down to $\sim 100 \, {\rm km \, s^{-1}}$ thanks to the VALES observations \citep{Villanueva2017arXiv170509826V}.  Our low--redshift SMGs have CO line width and luminosities overlapping with those for the high--redshift SMG population with CO detections, although the averages are lower for our low--redshift SMGs because at high--redshift, only the most luminous SMGs have $^{12}$CO detections (see Figure \ref{lumi_lumi_SMGs}).

\subsection{The dust--gas correlation}\label{section_dust_gas_correlation}

\begin{figure}[!t]
\centering
\includegraphics[width=0.45\textwidth]{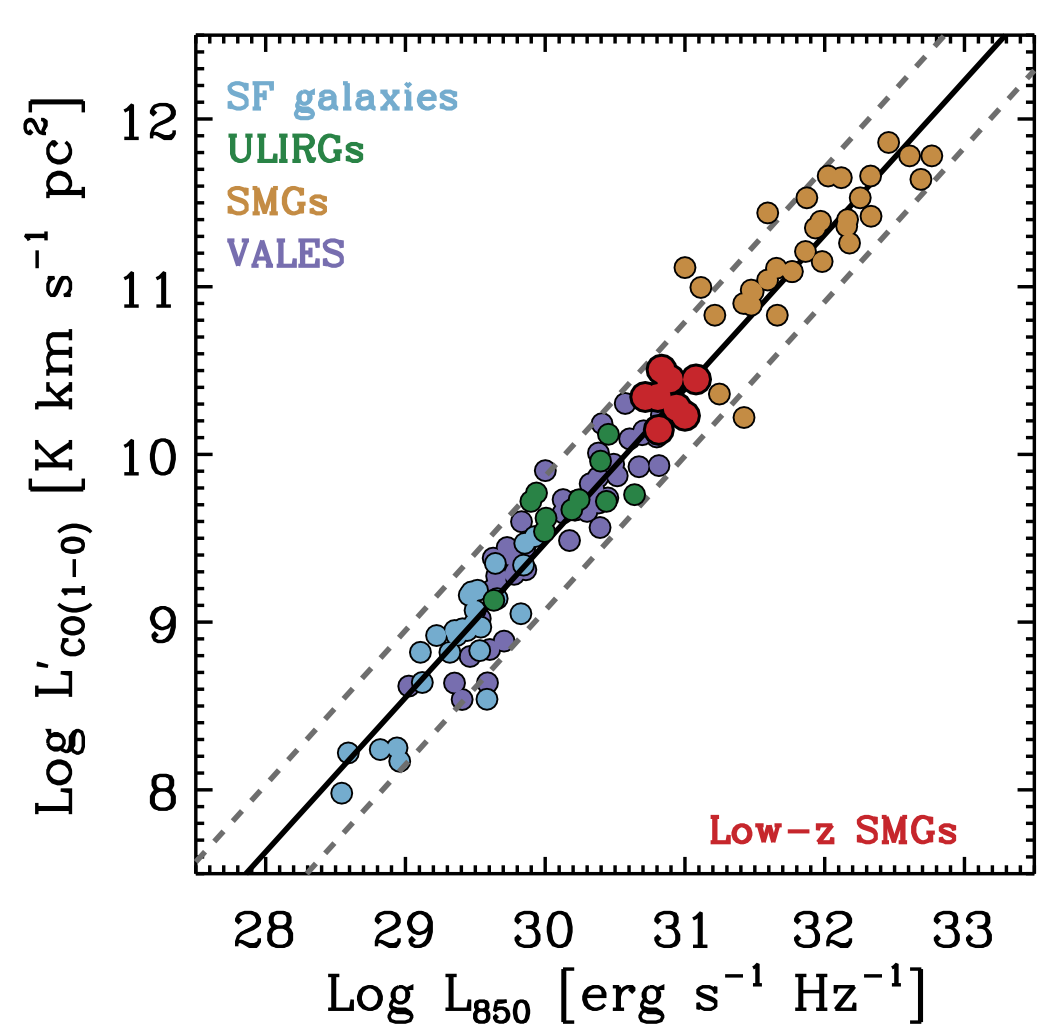} 
\caption{$^{12}$CO(1--0) luminosity versus luminosity at rest--frame $850 \, {\rm \mu m}$ for our low--redshift SMGs in comparison with other samples, including local star--forming galaxies, local ULIRGs and high--redshift SMGs \citep{Huynh2017MNRAS.467.1222H,Scoville2016ApJ...820...83S,Scoville2017ApJ...837..150S,Hughes2017arXiv170207350H,Fu2013Natur.498..338F,Ivison2013ApJ...772..137I,Harris2010ApJ...723.1139H,Thomson2012MNRAS.425.2203T,Greve2003ApJ...599..839G,Carilli2011ApJ...739L..33C,Ivison2011MNRAS.412.1913I,Aravena2013MNRAS.433..498A,Thomson2015MNRAS.448.1874T,Lestrade2011ApJ...739L..30L,Riechers2011ApJ...739L..31R,Harris2012ApJ...752..152H,Sanders1989A&A...213L...5S,Solomon1997ApJ...478..144S,Sanders1991ApJ...370..158S,Dale2012ApJ...745...95D,Young1995ApJS...98..219Y}.  The rest--frame $850 \, {\mu m}$ luminosities for our low--redshift SMGs have been determined from the MAGPHYS fits used to derive their stellar mass.  The same procedure has been used in VALES.  We see that our low--redshift SMGs nicely fall in the best-fitted relation derived from the VALES survey \citep{Hughes2017arXiv170207350H} shown with the black solid line (grey dashed lines represent $\pm 0.4 \, {\rm dex}$ the VALES relation), fills the gap between local and high--redshift galaxies, and supports the existence of a relatively tight correlation between the dust ($L_{\rm 850}$) and gas ($L'_{\rm CO(1-0)}$) luminosities for a wide range of redshifts and luminosities.
              }
\label{fig_lumi_L850}
\end{figure}

It has been reported in several previous works that there is a tight correlation between the luminosity at rest--frame $850 \, {\rm \mu m}$ and the $^{12}$CO(1--0) luminosity \citep{Dunne2000MNRAS.315..115D,Scoville2016ApJ...820...83S,Scoville2017ApJ...837..150S,Hughes2017arXiv170207350H}.  In this section we study the location of our low--redshift SMGs in the $L_{\rm 850} - L'_{\rm CO(1-0)}$ diagram.  This is represented in Figure \ref{fig_lumi_L850}, where we also include a compilation from previous work.  The $L_{850}$ luminosity for our low--redshift SMGs has been derived from the best--fit MAGPHYS templates used to estimate their stellar masses (see \S \ref{section_stellar_mass_GAMA}).

Our low--redshift SMGs are in excellent agreement with previously derived trends, and they start filling the gap between local and high--redshift galaxies.  This result supports the existence of a tight correlation (considerably tighter than the $L_{\rm IR} - L'_{\rm CO}$ correlation, see Figure \ref{lumi_lumi_SMGs}) between the dust ($L_{\rm 850}$) and gas ($L'_{\rm CO(1-0)}$) luminosities for a wide range of redshifts and luminosities.  This tight relation can be used to derive the $^{12}$CO(1--0) luminosity of sources in which that transition has not been observed and/or detected.  Actually, in the remaining sections of the paper we will use the gas--dust luminosity relation to measure the CO--related properties of the 13 low--redshift SMGs that have not been detected in $^{12}$CO(1--0), and also in those which have not been observed with the IRAM--30m telescope.


\subsection{Molecular and dust masses}\label{section_stellar_mass_GAMA}

\subsubsection{Molecular gas mass}

We showed in \S \ref{section_dust_gas_correlation} that our low--redshift SMGs with $^{12}$CO(1--0) emission follow very well the trend reported in previous work between the dust and gas luminosities.  We can then use that relation to estimate the molecular gas mass of all low--redshift SMGs, not only of those with detected $^{12}$CO(1--0) emission in our IRAM--30m observations.   Using an $\alpha_{\rm CO} = 0.8 M_\odot / \, {\rm K \, km \, s^{-1} \, pc^2}$ to convert from the $^{12}$CO(1--0) luminosity to the molecular gas mass, we obtain that the molecular gas mass of the full sample has an average of $M_{\rm H_2} \sim 1.6 \times 10^{10}\, M_\odot$, varying from $\sim 5 \times 10^{9} \, M_\odot$ to $\sim 3 \times 10^{10}\,M_\odot$. We then conclude that there is a significant variety on the molecular gas mass reservoirs, with almost an order of magnitude difference between the least and most massive sources. This is coupled with the large range in CO excitation that we found for our low--redshift SMGs in \S \ref{CO_SLED_section} and, in general, in agreement with the fact that low--redshift SMGs are a quite diverse population, similarly to what it has been claimed to happen at high--redshift \citep{Ivison2000MNRAS.315..209I}.

The most massive molecular gas reservoirs found in our low--redshift SMGs are similar to those found in the classical, high--redshift SMG population, although we find that many of our low--redshift SMGs are less massive in molecular gas.  Again, this is likely because, among the full sample of high--redshift SMGs with known spectroscopic redshifts, only the brightest have been observed in $^{12}$CO(1--0), which are also the most massive ones \citep{Ivison2011MNRAS.412.1913I,Riechers2011ApJ...739L..31R} according to the $L_{\rm IR} - L'_{\rm CO}$ correlation.

\subsubsection{Molecular gas mass fraction}\label{section_molecular_gas_fraction}

\begin{figure}[!t]
\centering
\includegraphics[width=0.45\textwidth]{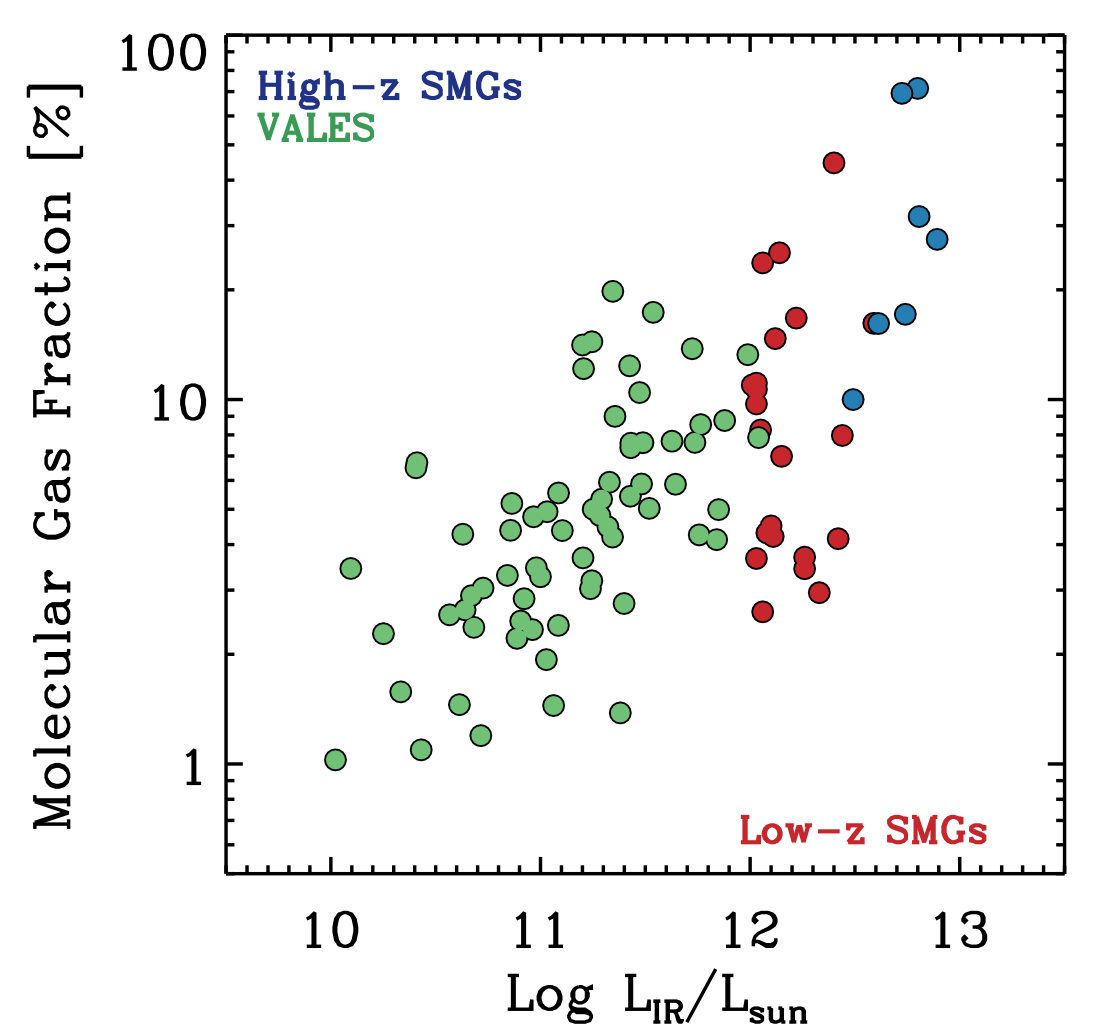} 
\caption{Molecular gas fraction, defined as $\mu_{\rm gas} = 100 \times M_{\rm gas} / (M_{\rm gas} + M_{\rm star})$, as a function of the total IR luminosity ($L_{\rm IR}$) for our sample of low--redshift SMGs.  We also include the galaxies in the VALES survey \citep{Villanueva2017arXiv170509826V}, which probe a significant range of total IR luminosities, and the high--redshift SMGs studied in \cite{Ivison2011MNRAS.412.1913I} and \cite{Huynh2017MNRAS.467.1222H}. All the stellar masses have been converted to Salpeter IMF.  We see that there is a trend, although with significant spread, where more IR--luminous galaxies have higher molecular gas fractions.  The spread of the correlation might be largely due to the uncertainties on the $\alpha_{\rm CO}$ factor (an $\alpha_{\rm CO} = 0.8 \, M_\odot / \, {\rm K \, km \, s^{-1} \, pc^2}$ has been used to convert the $^{12}$CO(1--0) luminosity into the molecular gas mass) and also the uncertainties in the determination of the stellar mass.  Even for our galaxies alone, which have a relatively narrow range of $L_{\rm IR}$, we see a large spread of molecular gas fraction, with the highest values being compatible with those found for bright SMGs with $^{12}$CO(1--0) detections.
              }
\label{fig_gas_fraction}
\end{figure}

We show in Figure \ref{fig_gas_fraction} the molecular gas fraction of our low--redshift SMGs, defined as $\mu_{\rm gas} = 100 \times M_{\rm gas} / (M_{\rm gas} + M_{\rm star})$, as a function of the total IR luminosity. Again, we see a quite significant variation, from galaxies with low ($\sim 2 \%$) to high ($\sim 50 \%$) molecular gas fraction. The highest molecular gas fractions found in our low--redshift SMGs are compatible with those found in high--redshift SMGs with $^{12}$CO(1--0) detections \citep{Ivison2011MNRAS.412.1913I,Riechers2011ApJ...739L..31R}. However, we should point out that only the most luminous high--redshift SMGs, those with $L_{\rm IR} > 10^{12.5} \, L_\odot$, have been detected in $^{12}$CO(1--0) and, therefore, these molecular gas fractions might be biased.  Furthermore, the sample studied in \cite{Ivison2011MNRAS.412.1913I} only includes SMGs whose $^{12}$CO(3--2) emission had been already detected in NOEMA, which might include an additional bias.  Actually, we can see in Figure \ref{selection_low_redshift_SMGs_TD_LIR} that there is a significant population of high--redshift SMGs with $12 < \log \left(L_{\rm IR} / L_\odot \right)< 12.5$, and none of them have $^{12}$CO(1--0) detections.  It might happen that those less luminous SMGs have lower molecular gas mass fractions, similar to those of most of our low--redshift SMGs. Actually, lower $^{12}$CO(1--0) luminosities would be expected for less luminous SMGs with $12 < \log \left(L_{\rm IR} / L_\odot \right)< 12.5$ if the $L_{\rm CO(1-0)} - L_{\rm 850}$ correlation holds for them.

We explore in Figure \ref{fig_gas_fraction} the relation between the molecular gas fraction and total IR luminosity by including the galaxies in the VALES survey \citep{Villanueva2017arXiv170509826V}.  We can see that there is a clear trend: more luminous galaxies tend to have higher molecular gas mass fractions.  However, the correlation has a huge scatter. Actually, as we discussed above, we see a significant range of molecular gas mass fractions in our low--redshift SMGs despite their total IR luminosities are within a relatively limited range. Part of this scatter might be real, but also part of it might be due to the number of uncertainties related to the estimation of the molecular gas mass, of which the $\alpha_{\rm CO}$ factor might be the most influential (note that here there is no uncertainty related to the CO line rations as all the measurements considered in this work have been obtained from the $^{12}$CO(1--0) emission).

\subsection{Star formation efficiency and gas depletion time}

We have found in previous sections that there is a noticeable variety in the molecular gas properties of our low--redshift SMGs. This is further confirmed when analyzing their star formation efficiencies, which can be parametrized by the continuum to line luminosity ratio $L_{\rm IR} / L'_{\rm CO}$. Our low--redshift SMGs with $^{12}$CO(1--0) detections have a wide range of $L_{\rm IR} / L'_{\rm CO}$ values, ranging within $40 < L_{\rm IR} / L'_{\rm CO} < 180$. The lowest values of the $L_{\rm IR} / L'_{\rm CO}$ ratio are compatible with those found for local disks, while the highest values are similar to those found in local ULIRGs and high--redshift SMGs \citep{Ivison2011MNRAS.412.1913I}. The lower limit on the $L_{\rm IR} / L'_{\rm CO}$ ratio for our low--redshift SMGs without $^{12}$CO(1--0) detections are $L_{\rm IR} / L'_{\rm CO} > 70$, and thus compatible with the values found for high--redshift SMGs.  Similar conclusions are obtained when considering the $^{12}$CO(1--0) luminosity of all 21 sources in our sample obtained from their dust continuum luminosity at rest--frame $850 \, {\rm \mu m}$, in which case the $L_{\rm IR} / L'_{\rm CO}$ values range within $38 < L_{\rm IR} / L'_{\rm CO} < 285$, with an average of $L_{\rm IR} / L'_{\rm CO} \sim 75$.

We obtain the gas depletion time of our low--redshift SMGs from the ratio between their molecular gas mass and their total SFR: $\tau_{\rm gas} = M_{\rm H_2} / {\rm SFR}$.  The derived values are within $15 \lesssim \tau_{\rm gas}\,[{\rm Myr}] \lesssim 120$, with an average of $\tau_{\rm gas} \sim 70 \, {\rm Myr}$. This means that our low--redshift SMGs will consume most of their gas in only a few tens of Myr, similar to what happens to SMGs at $z \sim 2$ and also to the most luminous galaxies in the early Universe \citep{Riechers2013Natur.496..329R,Ivison2013ApJ...772..137I,Oteo2016ApJ...827...34O,Riechers2017arXiv170509660R}.  After this time, it is likely that they become a massive, passively evolving system.

\subsubsection{Dust mass}

The dust mass of our low--redshift SMGs (see Table \ref{table_full_sample}) has been determined from the same MBB fits used to measure their dust temperature, and following equation 8 in \cite{Casey2012MNRAS.425.3094C} with a dust absorption coefficient of $\kappa_{850} = 0.15 \, {\rm m^2 \, kg^{-1}}$ \citep{Weingartner2001ApJ...548..296W,Dunne2003Natur.424..285D,Kovacs2010ApJ...717...29K}.  The assumed value for the dust absorption coefficient is the same as the used in high--redshift SMGs in the past \citep[see for example][]{Magnelli2012A&A...539A.155M,Swinbank2014MNRAS.438.1267S}. 

We see that the dust masses of our low--redshift SMGs vary within $0.5 \lesssim M_{\rm dust} \, [{\times 10^8 \, M_\odot}] \lesssim 2.3$, with an average value of $1.3 \times 10^8 \,M_\odot$.  This average value is slightly lower than some of those found in high--redshift SMGs \citep{Magnelli2012A&A...539A.155M,Swinbank2014MNRAS.438.1267S}, although the distributions still overlap.  It should be noted that the dust masses calculated in this work (and in most previous works) rely on the value assumed for the dust absorption coefficient. As a reference, this parameter in MAGPHYS is  $\kappa_{850} = 0.077 \, {\rm m^2 \, kg^{-1}}$, which is half the value assumed above and, consequently, the MAGPHYS--derived dust masses would be a factor of $\sim 2 \times$ higher than those derived from our MBB fits.

\subsubsection{Gas--to--dust ratio}

Using their gas and dust mass, we can now estimate the gas--to--dust ratio ($\delta_{\rm GDR}$) of our low--redshift SMGs, which ranges between $\sim 90$ and $\sim 170$, with an average value of $\sim 122$.  This average value is compatible (given the number of assumptions/uncertainties involved in its calculation) to that found in high--redshift SMGs, which is $\delta_{\rm GDR} \sim 90$ \citep{Magnelli2012A&A...539A.155M,Swinbank2014MNRAS.438.1267S}, and also compatible with those found in the Milky Way and local star--forming galaxies, being $\delta_{\rm GDR} \sim 130$ in both \citep{Jenkins2004oee..symp..336J,Draine2007ApJ...663..866D}.

\section{Conclusions}\label{section_conclusions_paper}

In this paper we have presented a sample of 21 low--redshift analogs of high--redshift SMGs selected from the {\it H}--ATLAS survey because their total IR luminosities and dust temperature match to those of the classical high--redshift SMG population.  As well as presenting the sample, we have also reported the molecular gas properties of 16 of those low--redshift SMGs obtained from observations taken with the IRAM--30m telescope.  Our main conclusions can be summarized as follows:

\begin{enumerate}

\item We find a diversity of morphologies in the near--IR imaging, from isolated sources resembling star--forming disks to systems in interaction or group of galaxies, with the latter two classes being formed by either several red sources or a combination of blue and red sources.  This represents the first evidence that low--redshift SMGs are a very diverse population (as it happens to their high--redshift counterparts -- \citealt{Ivison2000MNRAS.315..209I}), which is further confirmed by their variety in CO line ratios or molecular gas fractions.

\item The IRAM--30m observations revealed massive molecular gas reservoirs in our low--redshift SMGs, with masses ranging from $\sim 0.9$ to $\sim 2.5 \, \times 10^{10} \, M_\odot$ and an average of $\sim 1.6 \, \times 10^{10} \, M_\odot$.  Our low--redshift SMGs can be as massive as their high--redshift counterparts, although we note that only the most luminous SMGs have been imaged in $^{12}$CO(1--0), so the comparison is not completely fair.  $^{12}$CO(1--0) observations of less luminous high--redshift SMGs would provide a better comparison between both population, and also a better knowledge of the general SMG population.

\item We see a variety of line CO line profiles, from Gaussian to clearly double--peaked emission.  Among the four low--redshift SMGs with double--peak CO line profiles, two are relatively isolated sources (so the double peak line profile is likely associated to a rotating disk) and the other two show signs of interaction (so the double peak line profile is associated to a merger), according to their optical and near--IR imaging.  This suggests that caution must be taken when interpreting high--redshift SMGs with double--peak CO line profiles as mergers, and clearly highlights the interest of studying low--redshift SMGs to help interpret the properties of the high--redshift SMG population.

\item The CO line ratios in our low--redshift SMGs reveal a significant range of line excitations.  The main consequence of this is that extrapolations from $J > 1$ CO transitions to $^{12}$CO(1--0) can be highly uncertain and severely affect the estimation of the molecular gas mass from $J > 1$ CO emission in the absence of $^{12}$CO(1--0) detections, consistent with previous finding at high redshift.

\item The CO luminosities obtained for our low--redshift SMGs support the existence of a relatively tight correlation between the dust and gas luminosities (the $L'_{\rm CO} - L_{\rm 850}$ correlation), and a not--so--tight correlation between the total IR and CO luminosities.  Actually, our detections and upper limits in the $^{12}$CO(1--0) observations, in combination with data from the literature, reveal that for a fixed IR luminosity the CO luminosity can vary up to almost one order of magnitude.

\end{enumerate}


%

\begin{acknowledgements}
This work is based on observations carried out under project numbers 194-15 and 184-16 with the IRAM 30m telescope. IRAM is supported by INSU/CNRS (France), MPG (Germany) and IGN (Spain).  IO, RJI, LD, and ZYZ acknowledge support from the European Research Council in the form of the Advanced Investigator Programme, 321302, {\sc cosmicism}.  LD also acknowledges support from ERC Consolidator Grant, CosmicDust. IRS acknowledges support from STFC (ST/P000541/1), the ERC Advanced Investigator programme DUSTYGAL 321334 and a Royal Society/Wolfson Merit Award. TMH acknowledges the CONICYT/ALMA funding Program in Astronomy/PCI
Project N$^\circ$:31140020. DR acknowledges support from the National Science Foundation under grant number AST-1614213 to Cornell University. MJM acknowledges the support of the National Science Centre, Poland
through the POLONEZ grant 2015/19/P/ST9/04010. This project has
received funding from the European Union's Horizon 2020 research and
innovation programme under the Marie Sk{\l}odowska-Curie grant
agreement No. 665778. HD acknowledges financial support from the Spanish Ministry of Economy and Competitiveness (MINECO) under the 2014 Ramón y Cajal program MINECO RYC-2014-15686.  The {\it H}-ATLAS is a project with {\it Herschel}, which is an ESA space observatory with science instruments provided by European-led Principal Investigator consortia and with important participation from NASA. The {\it H}-ATLAS website is http://www.h-atlas.org/.
\end{acknowledgements}

\bibliographystyle{mn2e}

\bibliography{cii_references}

\appendix

\begin{figure*}
\centering
\includegraphics[width=0.80\textwidth]{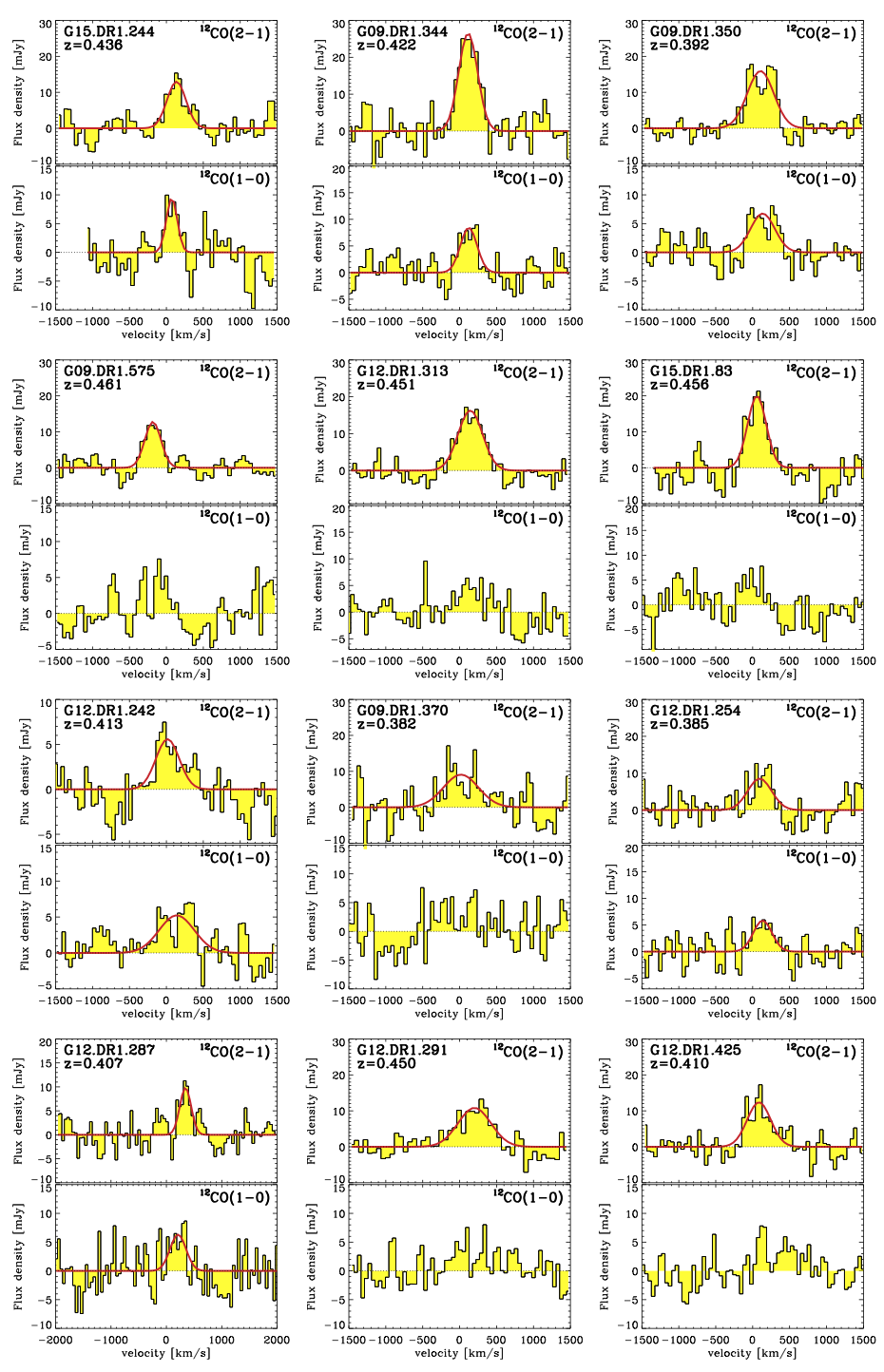}\\

\caption{CO spectra of our low--redshift SMGs observed with the IRAM--30m telescope.
              }
\label{CO_specs_fig_1}
\end{figure*}

\begin{figure*}
\centering

\includegraphics[width=0.80\textwidth]{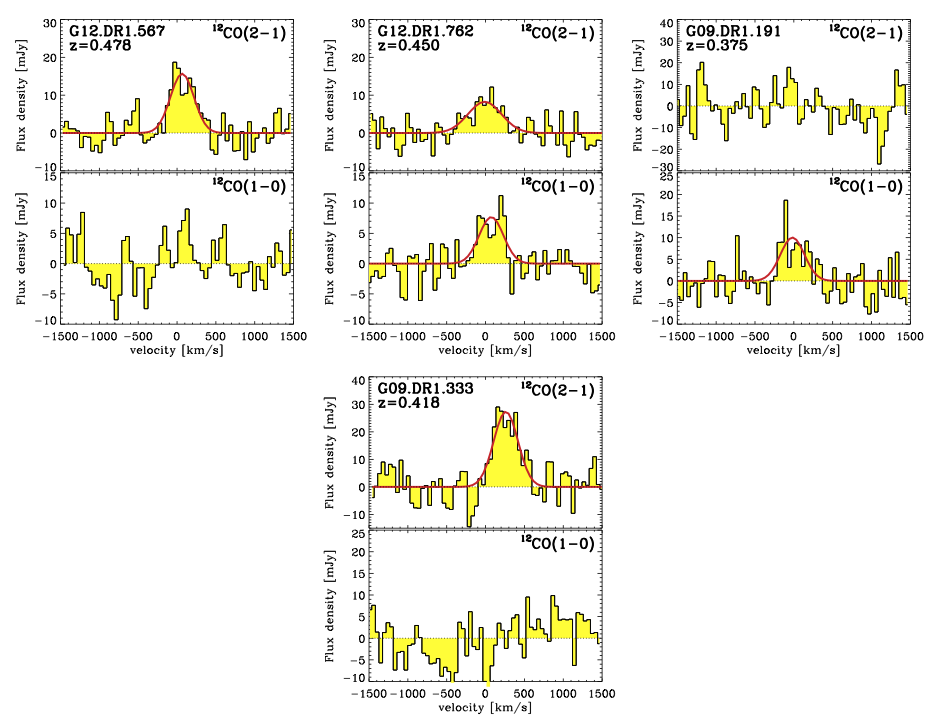}\\

\caption{CO spectra of our low--redshift SMGs observed with the IRAM--30m telescope ({\it Cont}).
              }
\label{CO_specs_fig_2}
\end{figure*}

\begin{figure*}
\centering
\includegraphics[width=0.80\textwidth]{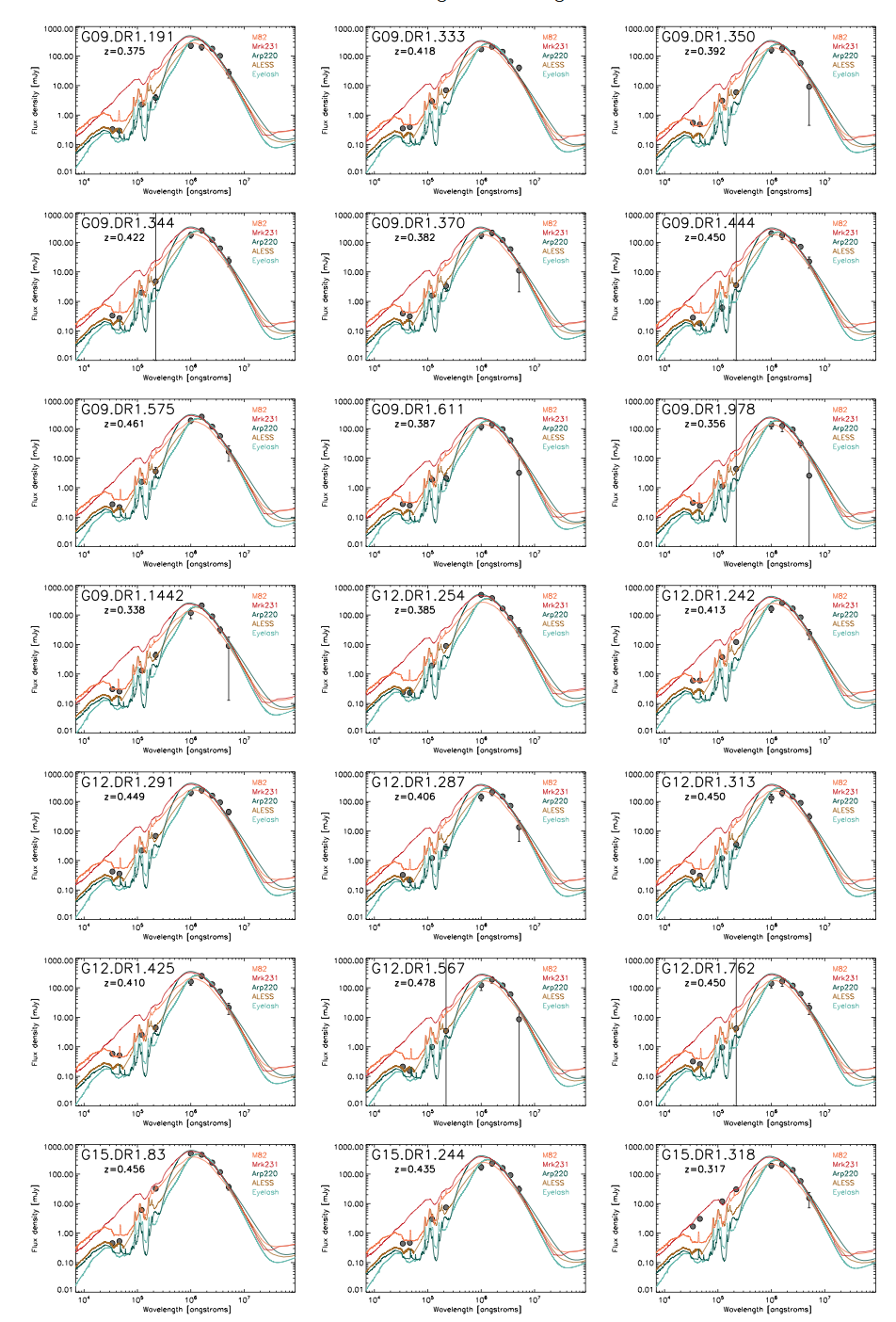} 

\caption{SED of the full sample of low--redshift SMGs studied in this work (see Figure \ref{sed_low_redshift_SGMS_WISE} for details)
              }
\label{sed_low_redshift_SGMS_WISE_appendix}
\end{figure*}

\end{document}